\documentclass[aps,groupedaddress, superscriptaddress,showpacs,amsmath,amssymb,twocolumn,prb]{revtex4-1}

\usepackage{bm}
\usepackage{graphicx}
\graphicspath{{figs/}}
\usepackage{wrapfig}
\usepackage{amssymb}
\usepackage{color}
\usepackage[normalem]{ulem} 
\usepackage{soul,xcolor} 
\setstcolor{black} 
\setul{}{2.4pt}
\usepackage{amsmath}
\usepackage{epstopdf}
\usepackage{mathtools} 
\usepackage{extarrows} 
\usepackage{bbm}
\usepackage{stmaryrd}
\usepackage{hhline}
\usepackage{amsfonts}
\usepackage{float}
\usepackage{hyperref}
\usepackage[caption=false]{subfig}
\usepackage{tikz}
\usepackage[siunitx, RPvoltages]{circuitikz}
\usepackage{comment}
\usetikzlibrary{calc}

\begin{document}

\title{Extracting Electromagnetic Bare Mode Couplings in Large Superconducting Quantum Processors}

\author{Reza Molavi}
\affiliation{Google Quantum AI, Santa Barbara, CA 93111, USA}
\author{Ebrahim Forati}
\affiliation{Google Quantum AI, Santa Barbara, CA 93111, USA}
\author{Yaxing Zhang}
\affiliation{Google Quantum AI, Santa Barbara, CA 93111, USA}
\author{Andrey R. Klots}
\affiliation{Google Quantum AI, Santa Barbara, CA 93111, USA}
\author{Juan Atalaya}
\affiliation{Google Quantum AI, Santa Barbara, CA 93111, USA}
\author{Brandon W. Langley}
\affiliation{Google Quantum AI, Santa Barbara, CA 93111, USA}
\author{Dogan A. Timucin}
\affiliation{Google Quantum AI, Santa Barbara, CA 93111, USA}
\author{Moein Nazari}
\affiliation{Google Quantum AI, Santa Barbara, CA 93111, USA}
\author{Ghazi Khan}
\affiliation{Google Quantum AI, Santa Barbara, CA 93111, USA}
\affiliation{Purdue Quantum Science and Engineering Institute, Purdue University, Indiana 47907, USA}
\author{Zlatko K. Minev}
\affiliation{Google Quantum AI, Santa Barbara, CA 93111, USA}
\author{Alexander N. Korotkov}
\affiliation{Google Quantum AI, Santa Barbara, CA 93111, USA}
\author{Michel H. Devoret}
\affiliation{Google Quantum AI, Santa Barbara, CA 93111, USA}
\affiliation{Department of Physics, University of California, Santa Barbara, CA 93106, USA}

\date{\today}

\newcommand{\AK}[1]{\textcolor{orange}{#1}}
\newcommand{\YZ}[1]{\textcolor{brown}{#1}}
\newcommand{\BL}[1]{\textcolor{blue}{#1}}
\newcommand{\ANK}[1]{\textcolor{blue}{#1}}
\newcommand{\DAT}[1]{\textcolor{red}{#1}}
\newcommand{\JA}[1]{\textcolor{green}{#1}}
\newcommand{\RM}[1]{\textcolor{olive}{#1}}
\newcommand{\JAsout}[1]{\textcolor{green}
{\sout{#1}}}

\begin{abstract}
High-fidelity control of superconducting quantum processors requires accurate characterization of electromagnetic coupling strengths among the device's constituent elements. Accurately extracting these couplings across large-scale architectures, presently featuring hundreds of qubits, poses a challenging multi-scale modeling problem. This requires resolving scales from the nanometer-scale geometry of Josephson junctions and their leads to the centimeter-scale size of the enclosing metallic packages. We present four numerical coupling extraction methods based on the avoided level crossing, the energy participation ratio, the induced electromotive force, and the impedance matrix. These methods are tailored to work with commercially available 3D electromagnetic solvers. We benchmark these techniques on a $10\times10$ array of transmon qubits, extracting their couplings to standing package modes. Our results show that these methods yield consistent coupling strengths with a maximum relative difference of less than 5\%.

\end{abstract}
\maketitle

\section{Introduction}

The hardware realization of a fault-tolerant quantum computer relies on the ability to engineer, characterize, and control coupled quantum systems with low error rates. Driven by these stringent requirements, monolithically integrated superconducting microwave circuits have established themselves as a promising hardware architecture. Most recently, scaling these planar architectures has enabled demonstrations of logical quantum error correction (QEC) operating below the surface code error threshold~\cite{google2023, google2025}. These scalable implementations predominantly utilize dense lattices of artificial planar atoms, most notably transmons~\cite{koch2007, houck2008, barends2013} in the circuit quantum electrodynamics architecture ~\cite{blais2004, blais2021}, interconnected via direct capacitive shunts or dedicated tunable coupling elements~\cite{chen2014, yan2018}.

In the state-of-the-art superconducting quantum processors, large-scale planar architectures are housed within relatively large metallic packages to isolate the system from environmental noise. Consequently, discrete electromagnetic modes (package modes) can appear within the qubit frequency band. These modes extend throughout the entire package volume and interact with the processor components. In particular, they can lead to elevated qubit decoherence and to long-range crosstalk errors, which can effectively reduce the code distance of surface codes and severely degrade QEC performance.

 \begin{figure}[htbp]
     \centering
     \includegraphics[trim=5cm 1.5cm 5cm -3cm, clip, width=0.95\linewidth]{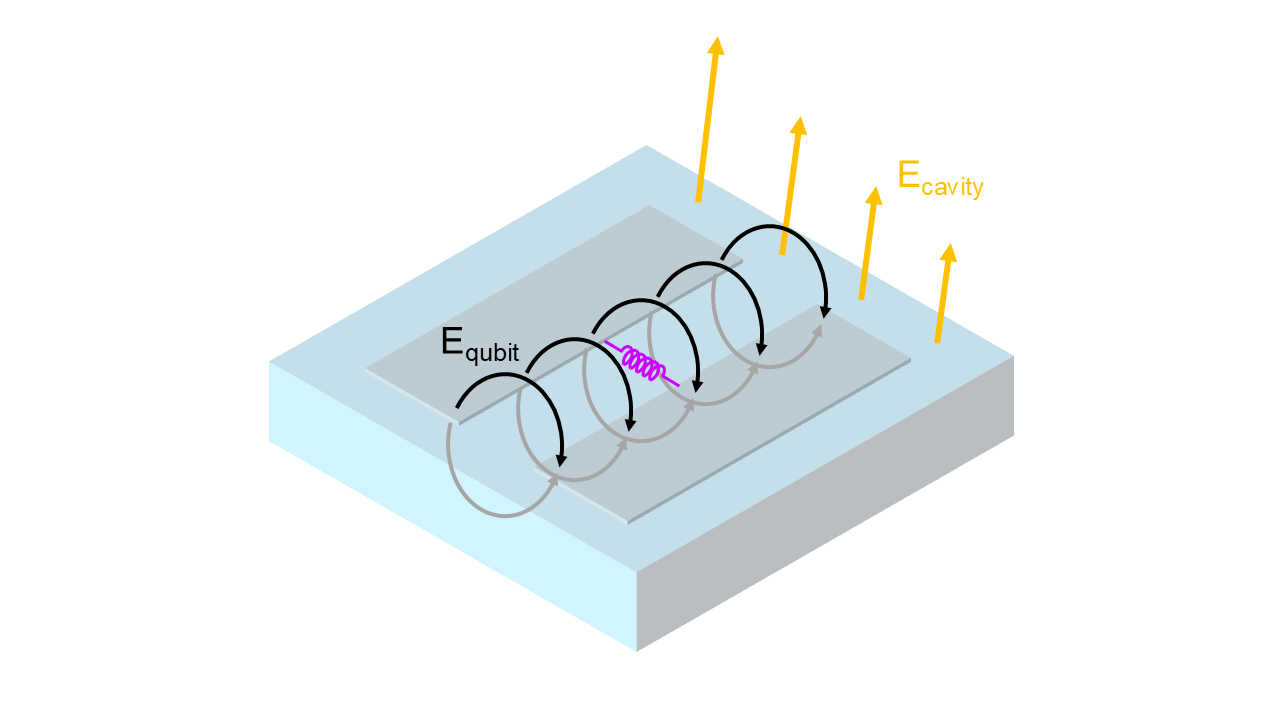}
 \caption{
 \textbf{Electric fields from transmon qubit and package mode.}
 The package mode electric field is non-uniform, mainly polarized along the vertical direction with a small lateral component. The qubit mode is differential, which means the total charge on both qubit paddles is zero. The blue box indicates a dielectric slab, and the coil (inductor) between the qubit paddles represents the Josephson junction. 
 }
 \label{fig:field_overlap}
\end{figure}

Designing the physical Hamiltonian for these processors relies on full-wave numerical electromagnetic calculations that demand strict precision. A critical challenge lies in extracting the minute couplings (with coupling efficiencies sometimes below $10^{-5}$) to the distributed package modes from the full electromagnetic solution. This presents a formidable multi-scale problem: models must account for the microscopic scale of the Josephson junction (JJ) and its leads while simultaneously capturing the full-wave dynamics of macroscopic box modes. When this spans dozens or hundreds of qubits, accurately modeling these disparate length scales—from nanometer-size components to centimeter-scale cavity boundaries—becomes a major computational bottleneck.

The qubit-package-mode coupling can be perturbatively calculated from the overlap integral~\cite{haus1991} of the electromagnetic fields of the qubit and package modes, see Fig.~\ref{fig:field_overlap}. However, this approach is computationally challenging  because it involves solving multiple independent geometries to extract the ``bare'' mode fields, and is highly sensitive to numerical errors when calculating fields across complex, multi-scale 3D domains~\cite{forati2026}.

Alternatives to explicit bare-mode field-overlap calculations use reduced circuit models, network response, or electromagnetic eigenmodes. Modular quasi-lumped modeling~\cite{minev2021lom} assembles individually modeled cells into interacting subsystems, accounting for loading and parameter renormalization, % ZKM: we could add that this methods wont work for the quantiies we want to compute here
while multiport impedance synthesis constructs equivalent networks from the electromagnetic response~\cite{solgun2015}. Black-box quantization~\cite{nigg2012} derives nonlinear Hamiltonians from the linearized circuit response. Energy-participation quantization~\cite{minev2021} obtains these Hamiltonians from modal energy participations without circuit synthesis, with subsequent work applying it to highly anharmonic circuits~\cite{yilmaz2026}. Explicit bare-mode parameters can also be recovered from participation information~\cite{yu2024}, while impedance-response and Green-function methods yield effective dispersive interactions~\cite{solgun2019,khan2024}. Full-wave spectral approaches further address mode hybridization and radiative loss in open structures~\cite{pham2025}.

%(ZKM: I tried to add an explicit challenge and motivation for the work; feel free to remove. It's optional, but at the moment I felt explicit over implicit is  helpful to the reader.) 
Despite these advances, accurately resolving weak qubit--package couplings in large processors remains challenging and computationally demanding: small local field amplitudes must be resolved within a multiscale electromagnetic structure, often in the presence of closely spaced resonances. This motivates a comparison and refinement of extraction methods with different numerical requirements.

\begin{figure*}[t] 
    \centering
    \includegraphics[width=\textwidth]{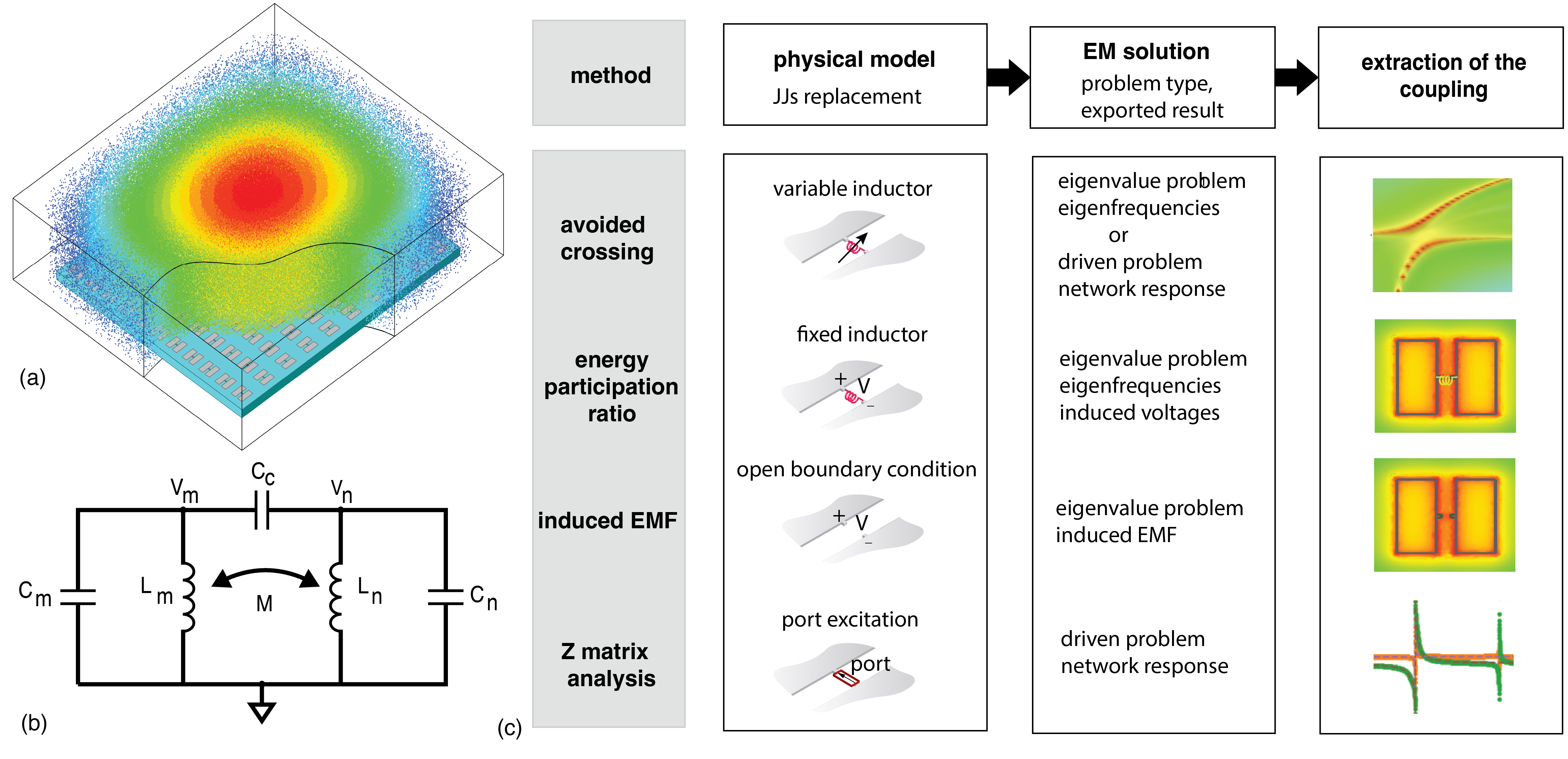}
    \caption{ 
    \textbf{Overview of the four coupling extraction methods.}
    (a) Full-wave electromagnetic model of a qubit array inside its metallic package, with a representative electromagnetic field distribution overlaid on the device geometry. 
    (b) Equivalent two-mode circuit illustrating the electric and magnetic coupling channels, represented by the coupling capacitance $C_c$ and mutual inductance $M$, respectively. 
    (c) The four methods are distinguished by the representation chosen for the Josephson junction (JJ) and by the electromagnetic quantity computed in the corresponding simulation. Avoided-crossing analysis replaces the JJ by a variable inductance and determines the total coupling $g$ from the eigenfrequencies, obtained either directly from an eigenmode calculation or from the driven network response. 
    The energy-participation-ratio (EPR) method uses a fixed junction inductance and determines $g$ from the fraction of modal energy stored in the JJ. 
    Induced-EMF analysis replaces the JJ by an open boundary and extracts $g$ from the voltage induced by the package mode. 
    Impedance-matrix analysis replaces the JJ by a lumped port and obtains $g$ from the driven network response.
    }    
    \label{fig:summary_of_methods}
\end{figure*}

In this work, we compare four methods for extracting couplings between localized qubit modes and distributed package modes: avoided crossing, energy participation ratio (EPR), induced electromotive force (EMF), and impedance matrix. The methods and their imposed boundary conditions are summarized in Fig.~\ref{fig:summary_of_methods}; the same techniques can also be used to compute qubit--qubit couplings. Each method uses numerical solutions of a {\it linear} electromagnetic model in which the JJs are replaced by method-specific boundary conditions. Qubits in this work are thus modeled as linear resonators. In these methods, the coupling strength is extracted from either the eigenfrequencies,  the eigenmode fields of the system, the induced EMF difference on the qubit ports, or the scattering matrix (S-parameters) of the microwave system as seen from the perspective of JJs. The S-parameters can alternatively be expressed as impedance or admittance matrices, thus we adopt the generalized term ``network response.''

The coupling extracted by all four methods can be understood 
from a common two-mode model. We therefore represent the qubit and the package mode as two coupled LC resonators $m$ and $n$ as shown in Fig.\ \ref{fig:summary_of_methods}(b). The linear coupling of two resonators  with bare frequencies $\omega_m$ and $\omega_n$ can be generally described by the Hamiltonian $H = \hbar\omega_m \hat a_m^\dagger \hat a_m + \hbar\omega_n \hat a_n^\dagger \hat a_n + H_{\rm int}$, where the last term describes the interaction,  

\begin{align}
    \frac{H_{\rm int}}{\hbar} &= g_C \frac{\hat{a}_m - \hat{a}_m^\dagger}{i}\frac{\hat{a}_n - \hat{a}_n^\dagger}{i} 
    - g_L(\hat{a}_m + \hat{a}_m^\dagger)(\hat{a}_n + \hat{a}_n^\dagger).
    \label{eq:hamiltonian}
\end{align}

Here $\hat a_m$ and $\hat a_m^\dagger$ are the bosonic ladder operators for resonator $m$, and likewise for resonator $n$. The coupling coefficients are $g_C$ and $g_L$; the negative sign for $g_L$ is because the flux-flux interaction energy is negative for positive mutual inductance.  It is customary to introduce the associated coupling efficiencies $k_C$ and $k_L$ via the relations  
\begin{equation}
    g_C=\frac{1}{2}k_C\sqrt{\omega_m \omega_n}, \,\,\,\,  g_L=\frac{1}{2}k_L\sqrt{\omega_m \omega_n},
    \label{eq:gs}
\end{equation}
 where, for the circuit in Fig.\ \ref{fig:summary_of_methods}(b), $k_C$  is the electric (capacitive) coupling efficiency proportional to the coupling capacitance $C_c$, while $k_L$  is the magnetic (inductive) coupling efficiency proportional to the mutual inductance $M$ (see Appendix~\ref{app:avoided_crossing}),   
\begin{equation}
    k_C = \frac{C_c}{\sqrt{(C_m+C_c)(C_n+C_c)}} , \,\,\,\,     k_L = \frac{M}{\sqrt{L_m L_n}} .
    \label{eq:ks}
\end{equation} When $\omega_m / \omega_n \approx 1$, we can use the rotating-wave approximation (RWA) which {enables us to drop terms of $H_{\rm int}$ that do not preserve the total number of excitations} (e.g., $\hat a_m \hat a_n$). Within the RWA, the total coupling strength $g$ is equal to   
\begin{equation}
    g = g_C - g_L,  
\label{eq:g-total}\end{equation}
which is the main object of interest that is computed by the four methods detailed below.

In the following sections, we first detail the conceptual foundations of the four methods. Next, we use a realistic example to explain the implementation details and compare their numerical results. Finally, we discuss the physics governing the qubit-package mode interactions.

\section{Methods descriptions} 

\subsection{Avoided-crossing method} \label{subsec:Avoided}
Avoided-crossing analysis is a conventional technique for extracting the total coupling $g$ between two resonators near resonance. In this method, we vary the JJ inductance ($L_{\rm qubit}$) in order to sweep the bare qubit mode frequency through the resonance of the bare package mode. Near resonance, the coupling strongly hybridizes the bare modes into symmetric-like and antisymmetric-like dressed eigenmodes with eigenfrequencies $\omega_\pm$, 
\begin{equation}
    \omega_\pm = \frac{\omega_m+\omega_n}{2} \pm \sqrt{\frac{(\omega_n-\omega_m)^2}{4} + g^2}.
    \label{eq:mode_pm}
\end{equation}
The absolute value of the total coupling strength is determined by the minimum difference of eigenfrequencies, attained when the bare modes are exactly on resonance ($\omega_m=\omega_n$); thus we have 
\begin{equation}
    |g| = \min(\omega_+ - \omega_-)/2.
    \label{eq:g_avoided_crossing}
\end{equation}

We can easily tune the qubit frequency during simulation by enforcing a variable inductive element in place of the JJ. This can be done in an electromagnetic eigensolver or, alternatively, in a circuit modeler (e.g., SPICE) after importing the network response (e.g., S-parameters) from a port-based (driven) electromagnetic solver. The latter approach is typically more accurate.

\subsection{EPR method}
\label{sec:EPR_method}
The EPR introduced in Ref.~\onlinecite{minev2021} measures the fraction of energy of the package eigenmode that resides inside the JJ of  a detuned qubit. This ratio depends on the physics of the interaction and allows us to extract the coupling strength between them. 

Specifically, the participation ratio $p_{mn}$ is defined as the ratio of the inductive energy $\mathcal{E}_{n,\, \text{ind}}$ stored in the JJ of qubit $n$ to the total \textit{inductive} energy of the package eigenmode $\mathcal{E}_{m,\, \text{ind}}$ (which constitutes half of the total mode energy $\mathcal{E}_m$), 
\begin{equation}
\label{eq:pmn}
    p_{mn}  \equiv \frac{\mathcal{E}_{n,\, \text{ind}}}{\mathcal{E}_{m, \,\rm ind}},  
     \quad \mathcal{E}_{n,\,\text{ind}} = \frac{1}{2} L_{n,\, \text{qubit}} \left( I_{n,\, \rm qubit}^{\rm rms} \right)^2.
\end{equation}
To evaluate the participation ratio, we measure the current $I_{n,\,\rm qubit}^{\text{rms}}$ through the lumped inductor that replaces the JJ, while energizing the package eigenmode with some arbitrary total energy $\mathcal{E}_m=2\mathcal{E}_{m, \,\rm ind}$. The JJ inductance $L_{n,\, \text{qubit}}$ is chosen such that the qubit and package  modes are sufficiently detuned to ensure a small participation ratio.

As derived in Appendix~\ref{app:epr}, the participation ratio $p_{mn}$ relates to the capacitive and inductive couplings as  
\begin{equation}
    \left| g_C - g_L \frac{\omega_m}{\omega_n} \right| = \frac{|\omega_m^2 - \omega_n^2|}{2\sqrt{\omega_m \omega_n}} \sqrt{p_{mn}}.
    \label{eq:g_tot}
\end{equation}
Using Eq.\ (\ref{eq:pmn}) and expressing the rms qubit current $I_{n,\, \rm qubit}^{\rm rms}$ via the voltage {\it amplitude} $V_{n,\, \rm qubit}$ across the JJ, we can rewrite this result as 
\begin{equation}
   g_C - g_L \frac{\omega_m}{\omega_n} =  \frac{\omega_m^2 - \omega_n^2}{2\sqrt{\omega_m \omega_n}} \, \frac{V_{n,\, \rm qubit}}{\omega_m\sqrt{2\mathcal{E}_m L_{n,\rm qubit}}}.
\label{eq:g_tot-2}\end{equation}

Equation~(9) gives the coupling combination $g_C - g_L(\omega_m/\omega_n)$ evaluated at the detuned qubit frequency.  As discussed below, this quantity can be converted to the sought on-resonance coupling $g$, so the condition $\omega_m/\omega_n \approx 1$  is not required. We reiterate that for the perturbative extraction used here, the modes should remain sufficiently detuned, $|\omega_m - \omega_n| \gg |g|$, so that the eigenmode remains predominantly package-like and its qubit participation is small.
Note that the relative sign between the voltage drops $V_{n,\, \rm qubit}$ across different qubits directly gives us the relative sign among $g$'s between the package mode $m$ and the various qubits in the array.

It may seem that, by changing the JJ inductance $L_{n,\rm qubit}$, we can distinguish the capacitive and inductive coupling components, $g_C$ and $g_L$. However, this is not the case, as both terms in the combination $g_C-g_L(\omega_m/\omega_n)$ scale as $\propto \sqrt{\omega_n}\propto L_{n,\, \rm qubit}^{-1/4}$ (assuming small magnetic coupling, so that $\omega_n \propto L_{n,\, \rm qubit}^{-1/2}$ -- see Appendix~\ref{app:avoided_crossing}). Therefore, 
\begin{equation}
     \left( g_C - g_L \frac{\omega_m}{\omega_n} \right) \sqrt{\frac{\omega_m}{\omega_n}} = g|_{\omega_n \to \omega_m}, 
\label{eq:g-res}\end{equation}
where $g|_{\omega_n \to \omega_m}\equiv g$ is the total coupling if the qubit were tuned on resonance with the mode $m$. Thus, the EPR method does not give us more information than the avoided-crossing method, except for the sign of $g$.

\subsection{Induced-EMF method}

In this approach, we replace the JJs with an open circuit in the electromagnetic simulator. We excite the package eigenmode with a specific energy $\mathcal{E}_m$ and measure the voltages induced across the qubits' open boundaries (ports). The total coupling between the qubit $n$ and the package mode $m$ can be related to the induced voltage {\it amplitude} $V_{n,\,\rm open}$ on the qubit port as 

\begin{equation}
    g  = \frac{\omega_m V_{n,\, \rm open}}{2} \sqrt{\frac{C_{n}+C_{c} }{2\mathcal{E}_m}}, \label{eq:v_o}
\end{equation}
where $C_{n}+C_{c}$ is the capacitance seen from $n$th qubit port (see Fig.\ \ref{fig:summary_of_methods}(b)) and  $\omega_m$ is the package mode frequency. Equation (\ref{eq:v_o}) can be derived directly from classical circuit theory  (see Appendix~\ref{app:induced_voltage}). Note that Eq.\ (\ref{eq:v_o}) can also be obtained from Eqs.\ (\ref{eq:g_tot-2}) and (\ref{eq:g-res}) by using the approximate formula $\omega_n=1/\sqrt{L_{n,\, \rm qubit} (C_n+C_c)}$ (for exact formula see Appendix \ref{app:avoided_crossing}) and taking the limit $L_{n,\, \rm qubit}\to \infty$.

This strategy becomes especially advantageous when calculating the coupling between a distributed package mode and each qubit of a large array. In the EPR method, the eigensolver must deal with a large number of dense qubit modes alongside the package modes. Conversely, in the induced-EMF analysis, replacing the JJs with an open circuit effectively pushes all qubit modes to zero frequency. The electromagnetic solver then only needs to target and resolve the few relevant package modes, thereby circumventing significant computational overhead. The benefit of the EPR method, however, is a more accurate high-frequency modeling of the qubit.

We calculate the induced voltage at each qubit port by integrating the electric field of the package mode along the shortest path connecting the corresponding JJ leads. This single calculation captures the induced voltage from both capacitive and  inductive couplings to the package mode, as detailed in Appendix~\ref{app:induced_voltage}. Similar to the EPR method, recording the sign of the simulated voltage drop ($V_{n,\,\text{open}}$) can provide us with the sign of the coupling $g$. 

Alternatively, the total coupling $g$ can be computed by imposing a short circuit instead of an open circuit at the JJs. As shown in Appendix~\ref{app:induced_voltage}, an equation analogous to Eq.~\eqref{eq:v_o} relates $g$ to the current flowing through the shorted qubit port.

\subsection{Impedance-matrix method}

The goal of the impedance-matrix analysis method is to construct an (approximate) lumped circuit model for the qubit-cavity system using the simulated impedance matrix obtained from the full electromagnetic solution. From the  circuit model, we then compute the coupling.

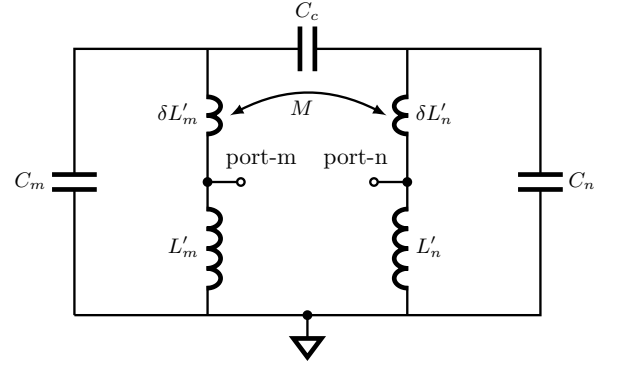
\begin{figure}[t]
\centering
\resizebox{0.9\linewidth}{!}{%
\begin{circuitikz}[american, line width=1pt]
\ctikzset{capacitors/scale=0.8,capacitors/thickness=2,inductors/thickness=2,grounds/thickness=2}
\ctikzset{bipoles/inductors/.cd, dot x distance=3pt, dot y distance=0pt}
\draw (0,0) to[C=$C_{{m}}$] (0,4)
to[C=$C_{{c}}$] (7,4)
to[C=$C_{{n}}$] (7,0)
to[short, -*] (3.5,0)
node[sground]{}
to[short] (0,0);
\draw (2,4) to[L, a=$\delta L'_{{m}}$, inductors/coils=2, inductors/width=0.35, name=l1, -*] (2,2)
to[short, -o] (2.5,2) node[right,font=\Large,yshift=10pt, xshift=-10pt]{$_{\textnormal{port-m}}$};
\draw (2,2) to[L, a=$L'_{{m}}$] (2,0);
\draw (5,4) to[L=$\delta L'_{{n}}$, mirror, inductors/coils=2, inductors/width=0.35, name=l2, -*] (5,2)
to[short, -o] (4.5,2) node[left,font=\Large,yshift=10pt,xshift=10pt]{$_{\textnormal{port-n}}$};
\draw (5,2) to[L=$L'_{{n}}$, mirror] (5,0);
\draw [latex-latex] ([xshift=0.05in,yshift=-0.0in]l1.north)
to[out=30, in=150] node[below, xshift=-0.1cm]{$M$}
([xshift=-0.15in,yshift=-0.0in]l2.south);
\end{circuitikz}%
}
\caption{\textbf{Equivalent lumped circuit model.}
The left and right ports represent the package-mode and qubit terminals, respectively. The inductive branches separate the port  inductances ($L'_m$ and $L'_n$) from the geometric inductances ($\delta L'_m$ and $\delta L'_n$) of the electromagnetic structure. The mutual inductance $M$ couples only geometric inductances. Together with the capacitances $C_m$ and $C_n$ and coupling capacitance $C_c$, these elements form the equivalent lumped  circuit whose impedance is fitted to the full-wave electromagnetic network response.
}
\label{fig:schematics_detailed}
\end{figure}

The equivalent lumped circuit model we use is shown in Fig.~\ref{fig:schematics_detailed}. Parameters $L^{\prime}_m$ and $L^{\prime}_n$ represent the lumped inductances  used to shunt the cavity and qubit ports (see Sec.~\ref{sec:Impedance_Matrix_Analysis}). Other circuit parameters are geometric capacitances ($C_m, C_n$, and $C_c$) and geometric inductances ($\delta L'_m$, $\delta L'_n$, \and $M$), extracted from the simulated impedance matrix. The coupling $g$ is given by Eqs.\ \eqref{eq:gs}, \eqref{eq:ks}, and \eqref{eq:g-res} with $L_m=L_m^{\prime}+\delta L_m^{\prime}$ and $L_n=L_n^{\prime}+\delta L_n^{\prime}$.

We determine the mode frequencies ($\omega_m$ and $\omega_n$) from the locations of poles in the simulated impedance matrix. Note that $\delta L'_n \ll L'_n$ ($L'_n$ is the JJ inductance, $\delta L'_n$ is the geometric inductance), while $\delta L'_m \gg L'_m$ ($L'_m$ is a small shunt, which should not disturb the package mode).

To extract the circuit parameters in Fig.~\ref{fig:schematics_detailed}, we fit the simulated impedance matrix as a function of drive frequency $\omega$ to the impedance matrix of the lumped circuit model, given by  
\begin{equation}
\mathbf{Z}\left(\omega\right)=\left(\frac{1}{j\omega\mathbf{L}^{\prime}}+\frac{1}{\left(j\omega\mathbf{C}\right)^{-1}+j\omega\delta\mathbf{L}^{\prime}}\right)^{-1},
    \label{eq:z_fit_new}
\end{equation}
where $j\equiv-i$ is the imaginary unit used in Electrical Engineering and matrices $\mathbf{L}^{\prime}$, $\delta\mathbf{L}^{\prime}$, and $\mathbf{C}$ are
\begin{align}
    &\mathbf{L}^{\prime} = \begin{pmatrix}  L_m^{\prime} & 0 \\ 0 & L_n^{\prime} \end{pmatrix}, \quad \delta\mathbf{L}^{\prime} = \begin{pmatrix}  \delta L_m^{\prime} &  M \\ M & \delta L_n^{\prime} \end{pmatrix}, \nonumber \\ 
    &\quad \mathbf{C} = \begin{pmatrix} C_m + C_c& -C_c \\ -C_c & C_n + C_c \end{pmatrix}.
\end{align}
Details of the fitting procedure are given in Appendix~\ref{app:lc_circuit}.  
While we focus on impedance matrix of a two-port network here, the method can be generalized to multi-port networks.

\section{Validation example and methods comparison}
\label{sec:simulation}

With the theory established, we put the four methods to the test on a sample quantum processor. We perform the full-wave electromagnetic simulations using the commercial electromagnetic solver~\cite{hfss}Ansys$^{\circledR}$ HFSS$^{\mathrm{TM}}$, utilizing both eigen- and driven-solver options, depending on the coupling extraction method.

\begin{figure}[t!]
    \centering
    \includegraphics[width=1.0\linewidth, trim= 0 0 0 1cm, clip=True]{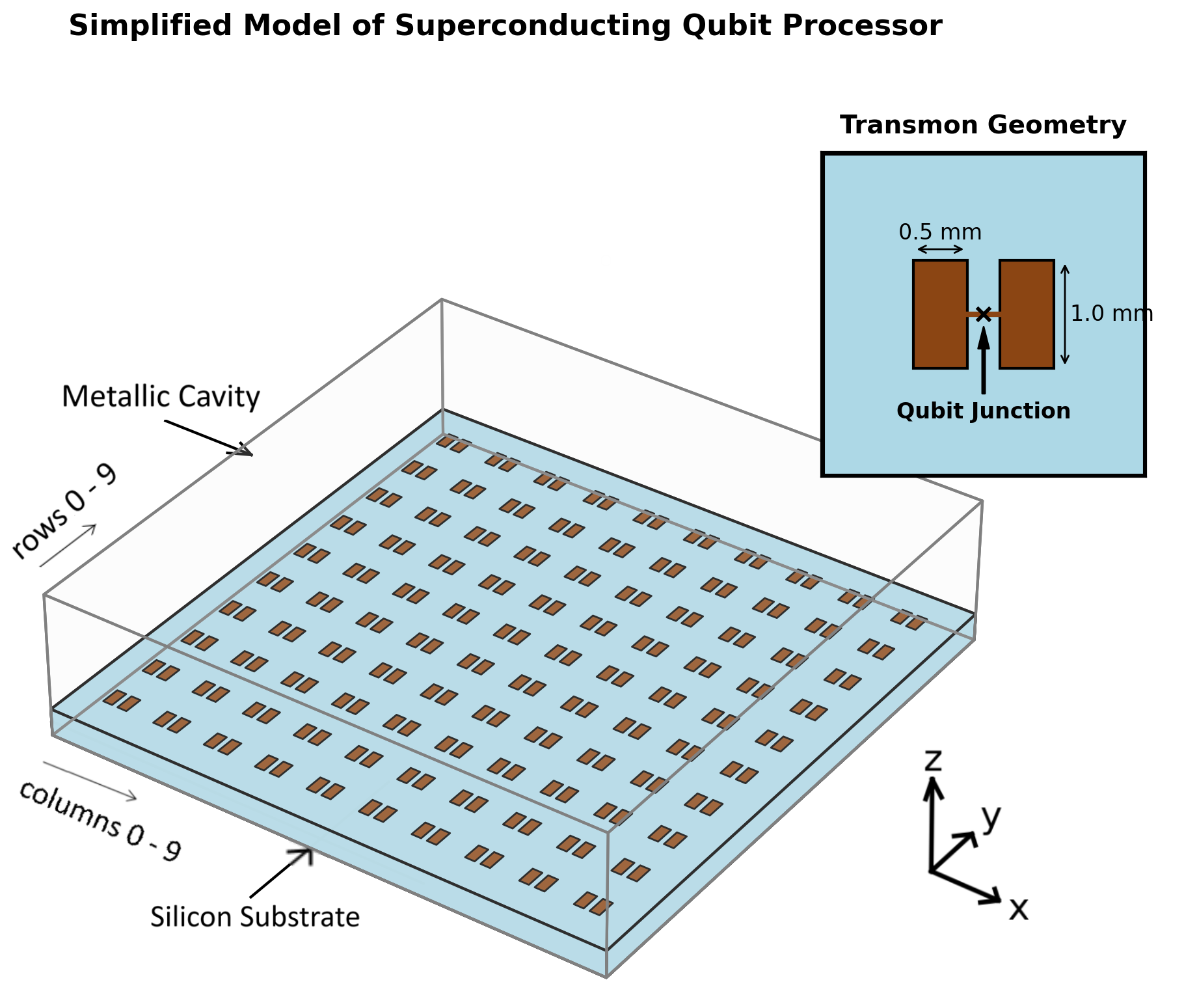}
    \caption{
    \textbf{Quantum processor for methods validation.} The simulated structure consists of a metallic box of size $30\,{\rm mm}\times30\,{\rm mm}\times3$~mm, partially filled with a Si substrate of 0.5~mm thickness. It includes a $10 \times 10$ array of planar transmons located on top of the substrate. The inset shows the transmon dimensions; the separation between the qubit paddle centers is 0.7~mm.}
    \label{fig:hfss}
\end{figure}

Figure~\ref{fig:hfss} displays the simulation geometry: a $10 \times 10$ array of   planar transmons of zero thickness with paddles of size 0.5~mm$~\times$~1~mm (separated by 0.2 mm), spaced with a period of 3~mm  in both $x$ and $y$ directions and housed inside a metallic cavity with dimensions $L_x=L_y=30$~mm and $L_z=3$~mm. The qubits sit on top of a silicon substrate with relative permittivity $\varepsilon_r = 11.9$ and thickness $t=0.5$~mm.

In this section, we numerically compute the (on-resonance) coupling $g$ of the qubits to the  fundamental package mode. In the absence of the dielectric and qubit paddles, this would be the conventional transverse-magnetic mode ${\rm TM}_{110}$, which has only a $z$-component of the electric field, $E_z^{\text{TM}_{110}} = E_0^{\rm TM} \sin ( \pi x/L_x ) \sin ( \pi y/L_y)$. The presence of the dielectric changes the mode, as discussed in Appendix~\ref{appendix:LSM_modes_derivation}.
The metallic qubit paddles further change the mode. In particular, without dielectric we would expect the fundamental mode frequency of 7.07 GHz; with the dielectric it becomes 6.49 GHz, and with additional qubit paddles (JJs are removed) it only changes to 6.48 GHz, indicating that the paddles almost do not change the eigenfrequency.

As we will see and discuss later, the spatial profile of the fundamental package mode coupling $g$ to different qubits is very different from the profile of $E_z^{\text{TM}_{110}}$ and instead is close to $\partial_x E_z^{\text{TM}_{110}}$. Note that the simulated structure does not have a ground-plane metallization, which is common in actual quantum processors. Therefore, we would expect relatively large couplings between qubits and the package mode, much larger than in actual processors.

\subsection{Avoided-crossing analysis}

Following  the methodology presented in Section~\ref{subsec:Avoided}, we replace the JJs with lumped ports in the electromagnetic solver and  replace one of the cavity sidewalls with a waveport in order to excite the fundamental package mode. The full-wave S-parameter data from the electromagnetic solver are embedded into circuit modelers such as SPICE or an alternative network analysis software such as scikit-rf Python package~\cite{scikit-rf}. We then terminate the cavity waveport to ground through a fixed $1\,\text{pH}$ inductance, which is sufficiently small to not perturb the package mode, but is still non-zero, so that we can calculate the impedance matrix and extract the eigenfrequencies from its poles. The qubit port is terminated with a variable inductance $L_{\rm qubit}$, which allows us to determine the system eigenfrequencies as a function of $L_{\rm qubit}$.

\begin{figure}[t]
        \centering
        \includegraphics[width=0.9\linewidth]{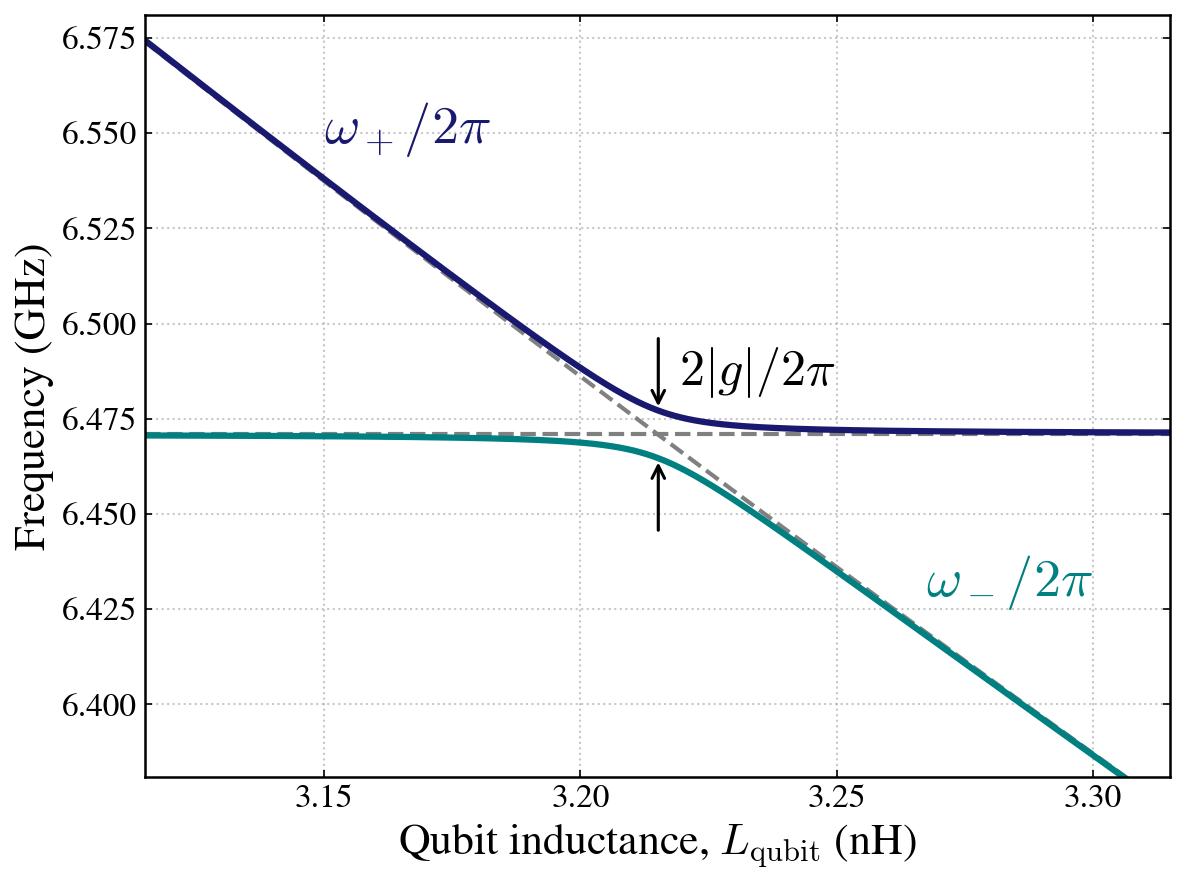}
        \caption{\textbf{Avoided-crossing spectrum for a representative transmon.} Sweeping the  inductance~$L_{\rm qubit}$ for qubit with array coordinates (1,1) (see Fig.~\ref{fig:hfss}) produces the minimum frequency separation $2 |g| = \min(\omega_+-\omega_-)$, giving us the coupling of this qubit to the package mode.}
        \label{fig:avoided_diagram}
    \end{figure}

Figure \ref{fig:avoided_diagram} shows an example of the eigenfrequency calculations for the qubit with array coordinates (1,1), with other qubits kept open. From the minimum separation between $\omega_+$ and $\omega_-$, we find $2|g|/2\pi =12.6$ MHz for the coupling between this qubit and the fundamental mode of the package. The procedure is automated for every  qubit in the array to map the spatial dependence of $|g|$ as a function of the qubits' position. The result is close to the absolute value of the result for $g$ shown in Fig.\ \ref{fig:g_avoided}, which is obtained by the induced-EMF method, discussed later.

\subsection{EPR analysis}

We use the HFSS eigenmode solver to compute the fundamental package eigenmode with one qubit at a time having the frequency of around 6.25 GHz ($\sim$0.2 GHz detuning) by choosing a proper $L_{\rm qubit}$, while other qubits are kept open. From the eigenmode, we compute the voltage drop $V_{\rm qubit}$ across $L_{\rm qubit}$ and then convert it to the coupling $g$ using Eqs.~\eqref{eq:g_tot-2} and \eqref{eq:g-res}. Note that computing the voltage drop $V_{\rm qubit}$ requires a precise line integration of the electric field between JJ leads. To ensure numerical accuracy, we utilize localized mesh seeding in the immediate vicinity of the integration line.

\begin{figure}[t]
    \centering
    \includegraphics[width=1.0\linewidth]{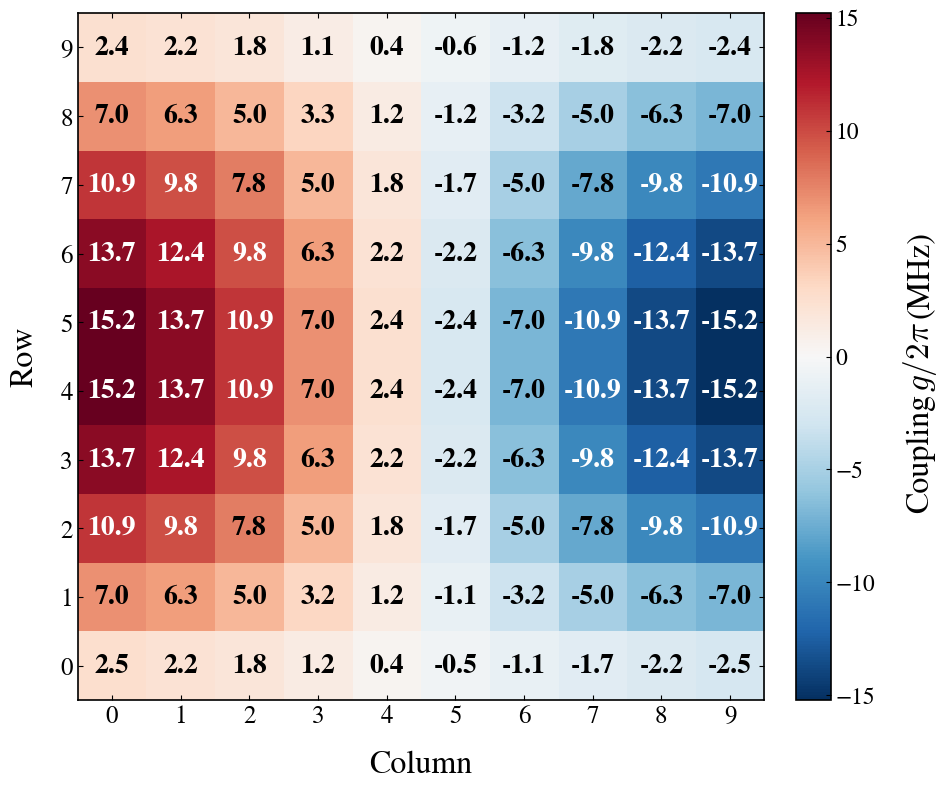}
    \caption{
    \textbf{Spatial profile of the couplings between qubits and fundamental package mode.}
      The couplings $g/2\pi$ are obtained using the induced-EMF method. The qubit coordinates are indicated by pairs of row and column indices ranging from 0 to 9. The spatial profile can be understood via Eq.~\eqref{eq:Ex_LSM-t}.     
    }
    \label{fig:g_avoided}
\end{figure}

The coupling $g$ to the package mode is computed for each qubit separately. The result is close to what is shown in Fig.\ \ref{fig:g_avoided}; comparison will be discussed later. Note that the sign of the coupling  $g$ is different for qubits with $x < L_x/2$ and $x > L_x/2$.

\subsection{Induced-EMF analysis}

This method achieves the most efficient computational parallelization for calculating the package-mode-to-qubit couplings, requiring the solution of just a single full-wave eigenmode simulation. By replacing the JJs with an open boundary circuit ($L_{\rm qubit} \rightarrow \infty$), the HFSS eigensolver computes only  the package mode of interest. With the induced voltages at the JJ ports obtained from the simulation, the couplings are then simultaneously extracted for  all qubits  using Eq.~\eqref{eq:v_o}. 

Figure~\ref{fig:g_avoided} depicts the spatial profile of the extracted couplings $g/2\pi$ between the fundamental package mode and the qubits.  
As mentioned above, the results of two previous methods are consistent with Fig.~\ref{fig:g_avoided}. We see that the couplings $g/2\pi$ are antisymmetric about the line $x=L_x/2$ and symmetric about the line $y=L_y/2$. A minor deviation from exact symmetry [e.g., compare qubits (0,0) and (9,0)] is due to the numerical inaccuracy. The difference between the behaviors along $x$ and $y$ directions is related to the geometry of the qubits (see Fig.\ \ref{fig:hfss}), for which qubit paddles are separated along $x$ direction. As mentioned above, relatively large couplings (on the order of 10 MHz) are due to the absence of the ground-plane metallization in the simulated structure.

\subsection{Impedance-matrix analysis}
\label{sec:Impedance_Matrix_Analysis}

\begin{figure*}[t!]
    \centering
    \includegraphics[width=\textwidth]{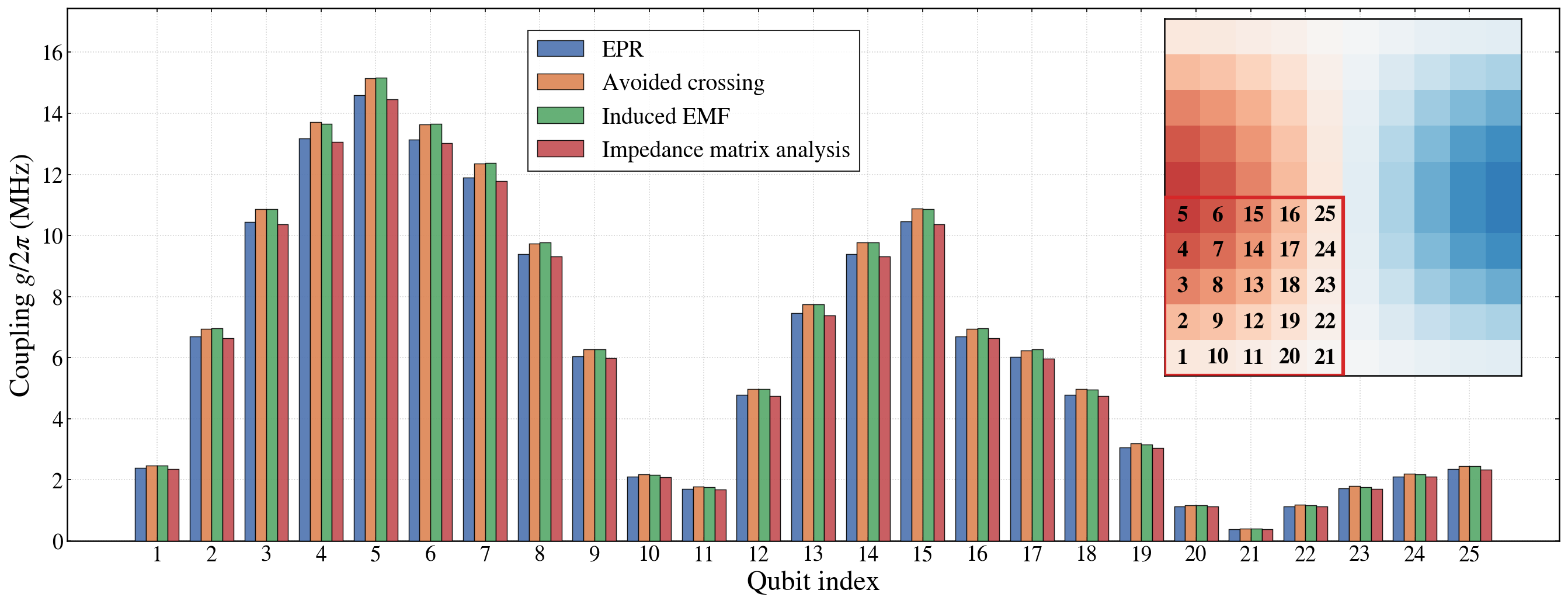}
    \caption{
    \textbf{Comparison of coupling extraction methods.}
    The couplings between the qubits and the fundamental package mode extracted via EPR, avoided-crossing, induced-EMF, and impedance-matrix methods are shown for the 25 qubits, with the numbering indicated in the inset. Because of the coupling symmetry (see the inset and Fig.\ \ref{fig:g_avoided}), data is presented only for one quadrant of the $10\times10$ qubit array. The four methods yield couplings with a maximum relative difference of only 4.8\% across the entire qubit array.
    }
    \label{fig:compare}
\end{figure*}

The simulation setup of the impedance-matrix analysis method is similar to the avoided-crossing method setup.
In the electromagnetic solver, we replace JJs with lumped ports and also replace one of the cavity sidewalls with a waveport to efficiently excite the intended  package mode. We then perform a driven network simulation to obtain the simulated impedance matrix as a function of drive frequency $\omega$. 
In the post-processing, we terminate the qubit port of interest with a fixed lumped inductance $L_{\rm qubit}$ leaving other qubits open and terminate the cavity port with a small inductance $L_{\rm cavity}=1$~pH. We then truncate the resulting impedance matrix (now including the shunting inductances) to a two-by-two matrix containing the ports for the cavity and qubit of interest.
In contrast to the avoided-crossing method, we do not sweep the qubit shunting inductance $L_{\rm qubit}$, but analyze the impedance matrix as a function of $\omega$ to extract the circuit parameters. Specifically, we choose $L_{\rm qubit} = 3.6$~nH (so that the qubit frequency is detuned by $0.39$~GHz below the fundamental package mode) and then use the simulated impedance matrix computed for the frequency range between 6~GHz and 7~GHz with 0.1~MHz resolution; this covers the frequencies of both the qubit and the package mode.

Fitting the simulated impedance matrix to Eq.~(\ref{eq:z_fit_new}) allows us to extract all the circuit parameters in Fig.~\ref{fig:schematics_detailed} and then compute the coupling $g$ via Eqs.\ (\ref{eq:gs}), (\ref{eq:ks}), and (\ref{eq:g-res}) -- see Appendix~\ref{app:lc_circuit}. As an example, let us discuss results for the qubit with array coordinates (0,0).  
First, we find that the extracted port inductances $L'_n=3.63~{\rm nH}$ and $L'_m=1~{\rm pH}$ closely match the introduced shunting inductances, thus validating the fitting procedure. 
Next, for the qubit we extract the capacitance $C_n = 173.8$~fF and geometric inductance $\delta L'_n = 0.32$~nH. Both values agree well with the independently simulated values from electrostatic and magnetostatic simulations. The remaining extracted values are $C_m= 132.0$ fF, $C_c= 100$ aF, $\delta L'_m= 4.57$ nH, and $M= -0.34$ pH. From this data we find $g_C/2\pi=  2.07$ MHz and $g_L/2\pi= -0.25$ MHz, so that $g/2\pi= 2.42$ MHz, according to Eq.~\eqref{eq:g-res}. This procedure is automated for all qubits in the array. The resulting couplings $g$ are close to what is shown in Fig.~\ref{fig:g_avoided}. 

It may seem that in contrast to other methods, this method can distinguish the capacitive and inductive contributions $g_C$ and $g_L$. However, this is not the case. In particular, if we change the location of the cavity port from the sidewall along $x$ to the sidewall along $y$ direction, then the values of $g_C$ and $g_L$ change significantly, while the value of $g$ remain the same. The impossibility to distinguish $g_C$ and $g_L$ by this method is expected because of the distributed nature of the package mode.

\subsection{Comparison of the methods}

All four methods produce the results for the coupling $g$ close to what is shown in Fig.\ ~\ref{fig:g_avoided} (except that the avoided crossing method can only produce $|g|$). For a detailed comparison, we show the results for all methods side-by-side in Fig.~\ref{fig:compare}. Since the symmetry about the line $y=L_y/2$ and antisymmetry about $x=L_x/2$ is present in all the methods, we show only the left bottom quadrant. The numbering for 25 qubits in the array is displayed in the inset. 

Most importantly, we see that all four methods produce almost equal results for $g$. The maximum relative difference is 4.8\%, the average relative difference is 2.9\%. The avoided-crossing and EMF methods tend to produce slightly bigger coupling compared to the EPR and impedance matrix methods. 

The presented cross-validation of the methods means that the choice of a method is a matter of convenience. From the point of view of the simulation time, the most convenient method is based on the induced EMF; it allows us to compute couplings to all qubits from only one full-wave eigenmode simulation, though it also requires knowing the capacitances for all qubits. However, at least in some situations, the EPR-based method can be more accurate, since its full-wave eigenmode simulation treats the qubit current explicitly. Moreover, by using sufficient detunings in the EPR method, we can also extract couplings to many qubits at the same time (in contrast to one-by-one extraction used in this paper). On the other hand, the methods based on avoided crossing and impedance matrix can be more accurate in some situations because they use driven full-wave simulations, which are typically more accurate than the eigenmode simulations.

\section{Physics of qubit couplings to the package mode} 

\begin{figure}[t]
    \centering
    \includegraphics[width=1.0\linewidth]{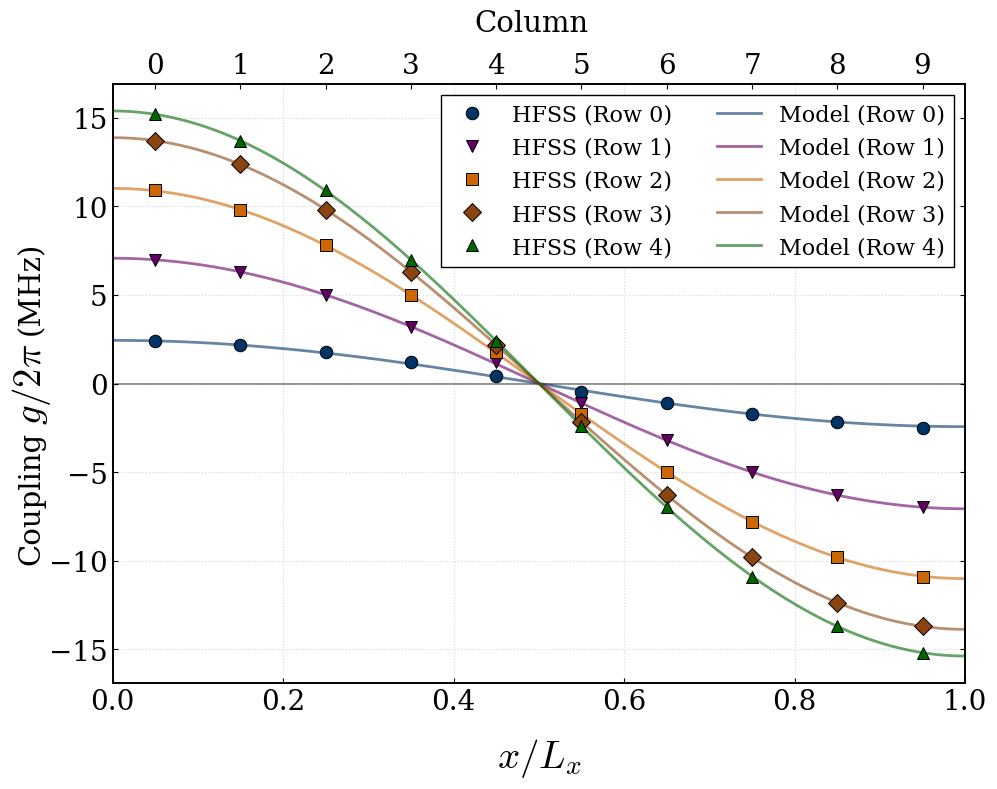}
    \caption{
    \textbf{Qubit couplings to the fundamental package mode.} Symbols show the qubit couplings $g/2\pi$ for the first five rows of the array (same data as in Fig.\ \ref{fig:g_avoided}) versus the $x$ coordinate of the qubit center. The solid lines are fits with the function $A \cos(\pi x/L_x)\sin(\pi y/L_y)$, where $y$ is the $y$-coordinate of the qubit center in a row and  $A=15.6$ MHz is the only fitting parameter. 
    }
    \label{fig:fit_fund}
\end{figure}

% \ANK{\bf Sasha's version}
To understand physics of the simulated qubit couplings to the fundamental mode of the package, let us draw the data shown in Fig.\ \ref{fig:g_avoided} in a different way -- see Fig.\ \ref{fig:fit_fund}. Here different symbols correspond to different rows in Fig.\ \ref{fig:g_avoided} (from row 0 to row 4), vertical axis is the extracted coupling $g/2\pi$, and the horizontal axis is the ratio $x/L_x$ for the $x$-coordinate of the \textit{qubit center}, i.e. junction location inside the grid -- see Fig.\ \ref{fig:hfss}, within the package length, $0< x<L_x$. The lines are the fit of the data to the dependence 
  \begin{equation} 
g=A \cos (\pi x/L_x)\sin (\pi y/L_y),
    \label{eq:g-profile}\end{equation} 
with $y$ being the $y$-coordinate of the qubit center, $0< y<L_y$. As we see, there is an excellent fit for all qubits with the single fitting parameter $A/2\pi = 15.6$~MHz. Therefore, our understanding of the coupling physics should explain the dependence (\ref{eq:g-profile}). 

Note that this $g$-profile is very different from the electric field profile of the fundamental ${\rm TM}_{110}$ mode,  $E_z^{\text{TM}_{110}} = E_0^{\rm TM} \sin ( \pi x/L_x) \sin ( \pi y/L_y)$. However, the $g$-profile (\ref{eq:g-profile})  is proportional to its $x$-derivative, $\partial_x E_z^{\text{TM}_{110}}$, with the asymmetry between $x$ and $y$ related to the orientation of qubits inside the cavity-- see Fig. \ref{fig:hfss}. 

It is possible to understand the coupling profile  (\ref{eq:g-profile}) in the following way. The presence of the dielectric substrate at $0<z<t$ significantly modifies the fundamental  ${\rm TM}_{110}$ mode. We will name this modified mode as ${\rm LSM}_{110}$, following the terminology of Longitudinal-Section-Magnetic (LSM) modes\cite{balanis2012}. More generally, let us discuss the mode ${\rm LSM}_{ab0}$, which is the  modification of ${\rm TM}_{ab0}$. This mode is  considered in detail in Appendix \ref{appendix:LSM_modes_derivation}. 
The $z$-component of its electric field,  $E_z^{\text{LSM}_{ab0}}$, jumps 11.9 times at the dielectric surface (due to silicon permittivity) and has a weak $z$-dependence away from the surface. 
Most importantly, this modified mode has also a non-zero $x$-component of the electric field at the dielectric surface ($z=t$),   
\begin{equation}
    E_{x\,\,\, |z=t}^{\text{LSM}_{ab0}} = a E_{0,x} \cos (a\pi x/L_x) \sin ( b \pi y /L_y),
    \label{eq:Ex_LSM-t}
\end{equation} 
where $E_{0,x}$ is a normalization constant ($E_{0,x}$ almost does not depend on $a$ and $b$ if we normalize the mode energy). The qubit has a non-zero electric dipole moment in the $x$ direction which interacts with $E_x$,  thus this qualitatively explains the coupling profile (\ref{eq:g-profile}), which has the the same dependence on $x$ and $y$ as Eq.\ (\ref{eq:Ex_LSM-t}) for $a=b=1$. Note that the mode ${\rm LSM}_{ab0}$ also has a non-zero $y$-component of the electric field; however, it is not important for us since the $y$-component of the qubit dipole moment is zero because of the symmetry of the qubit charge distribution.

Now let us estimate the coefficient $A$ in Eq.\ (\ref{eq:g-profile}) using the electric dipole interaction model. We will associate this coupling with $g_C$ and fully neglect the magnetic coupling $g_L$ (we assume $g_L$ is small because typically only $\sim 4\%$ of qubit energy is in the magnetic field). 
Relating $g_C$ to the coupling efficiency $k_C$ via Eq.\ (\ref{eq:gs}), we estimate $k_C$ for a qubit with center  coordinates $(x,y)$ as 
\begin{equation}
    k_C(x,y) = \frac{p_x E_{x\,\,\, |z=t}^{\text{LSM}_{ab0}} (x,y)}{2 \sqrt{\mathcal{E}_m \mathcal{E}_{n,\, \rm qubit}} }, 
    \label{eq:k_dipolar_coupling}
\end{equation}
where $p_x$ is the electric dipole moment in $x$ direction for the qubit mode with energy $\mathcal{E}_{n,\, \rm qubit}$, while $E_{x\,\,\, |z=t}^{\text{LSM}_{ab0}} (x,y)$ is the position-dependent $x$-component of the electric field of the LSM mode with energy $\mathcal{E}_m$. The product $p_x E_x$ in the numerator can be interpreted as the work performed by the electric field $E_x$ on the qubit charge $q$ moved by a distance $\Delta x$ (from paddle to paddle), so that $p_x=q \Delta x$. The factor of 2 in the denominator is because both $p_x$ and $E_{x\,\,\, |z=t}^{\text{LSM}_{ab0}}$ denote the amplitude values of the oscillations, while we need rms values for the average interaction energy (we reiterate that the magnetic-field interaction is neglected). 

It is customary to use 1 Joule of energy for the electromagnetic modes in the analysis, $\mathcal{E}_m=\mathcal{E}_{n,\, \rm qubit}=1\,{\rm J}$. Using the HFSS eigenmode solver, we compute $p_x$ by evaluating the surface integral~\cite{jackson1999} $p_x = \int_{\mathcal{S}_{\rm qubit}}\text{d}x \text{d}y \, J_x(x,y) / (j\omega_{\rm qubit})$, where $J_x(x,y)$ is the surface current density of the qubit, integrated over its surface $\mathcal{S}_{\rm qubit}$ (working with the current density in HFSS is more straightforward than working with the charge density). This gives us $p_x \approx 3.41 \times 10^{-10}~{\rm C\cdot m}$. Note that the corresponding  dipole length is $\Delta x = p_x/\sqrt{1\,{\rm J}\times 2C_{\rm qubit}}\approx 0.58$~mm, where $C_{\rm qubit} \approx 170~{\rm fF}$. This length is somewhat smaller than the 0.7~mm distance between the qubit paddle centers, as expected from the concentration of charges near the inner edges of the two paddles for the differential qubit mode. To calculate $E_{x\,\,\, |z=t}^{\text{LSM}_{110}} (x,y)$ for the fundamental LSM mode, we follow the procedure outlined in  Appendix~\ref{appendix:LSM_modes_derivation}, and for 1 Joule of energy obtain $E_{0,x} = 25.56\times 10^6 \,{\rm V/m}$ in Eq.\ (\ref{eq:Ex_LSM-t}). 

Combining these results for $p_x$ and $E_{0,x}$, we obtain the coupling efficiency $k_C(x,y)=4.36\times 10^{-3} \cos (\pi x/L_x) \sin (\pi y /L_y)$. Correspondingly, we explain the numerically obtained coupling profile in Eq.\ (\ref{eq:g-profile}) and estimate the parameter $A$ to be $A/2\pi = 14.2$ MHz. This agrees sufficiently well with  the numerical value of $A/2\pi = 15.6$, thus validating the electric-dipole-interaction approach. Note that we fully neglected the small magnetic interaction, which can be responsible for the small difference.  

\begin{figure}[t]
    \centering
    \includegraphics[width=1.0\linewidth]{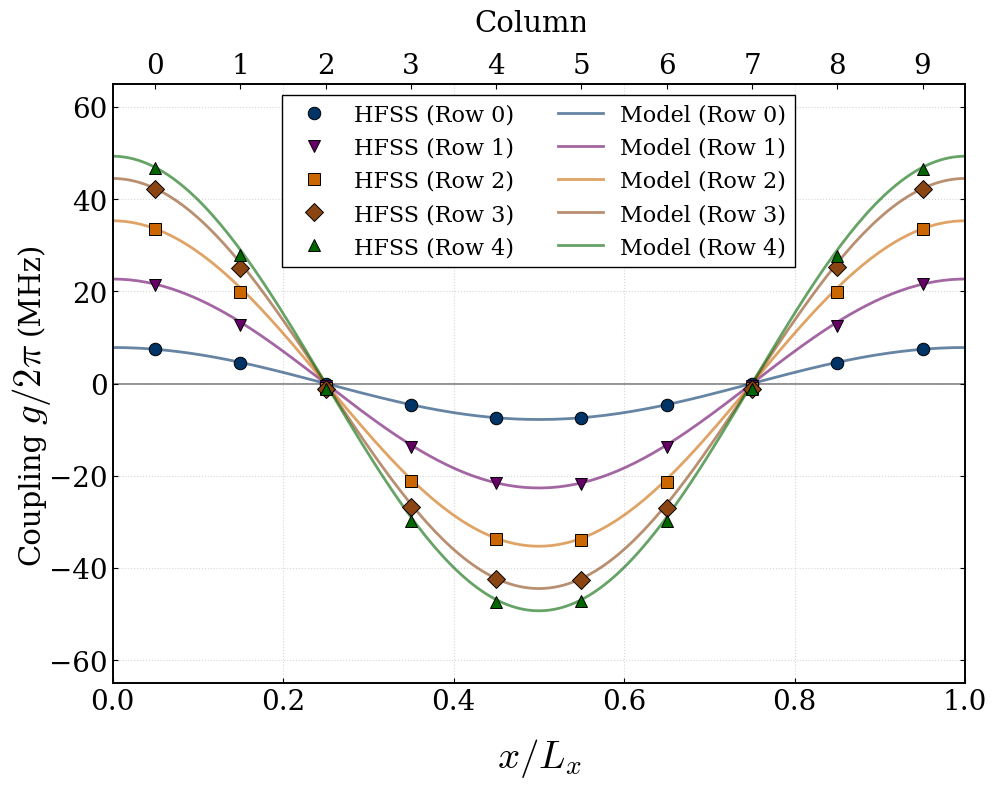}
    \caption{
    \textbf{Qubit couplings to the package mode $\text{LSM}_{210}$.} Similarly to Fig.\ \ref{fig:fit_fund}, symbols show the numerically extracted couplings of the qubits in the first five rows of the array. The solid lines are fits by the function $A_{210} \cos(2\pi x/L_x)\sin(\pi y/L_y)$, with the single fitting parameter $A_{210}=49.8$ MHz.
        }
    \label{fig:fit_second}
\end{figure}

To further validate the model, let us consider the qubit couplings to the second package mode, ${\text{LSM}_{210}}$, which has the frequency of $\omega/2\pi =10.23$ GHz (without qubit paddles).   
We can still use Eq.\ (\ref{eq:Ex_LSM-t}) with the slightly different value of $E_{0,x}$ for 1 joule of energy: $E_{0,x} = 26.25\times 10^6 \,{\rm V/m}$. This gives us $k_C(x,y)=2\times 4.47\times 10^{-3} \cos (2\pi x/L_x) \sin (\pi y/L_y)$  and  therefore the model predicts the on-resonance couplings 
  \begin{equation} 
g_{210}=A_{210} \cos (2\pi x/L_x)\sin (\pi y/L_y),
    \label{eq:g-profile-2}\end{equation} 
with the estimated value $A_{210}/2\pi =45.7$ MHz. The numerical results from the induced-EMF method are presented in Fig.\ \ref{fig:fit_second}. We see that the coupling profile indeed agrees with Eq.\ (\ref{eq:g-profile-2}), with a slightly different maximum value, $A_{210}/2\pi = 49.8$ MHz. Note that the frequencies of the modes ${\text{LSM}_{210}}$ and ${\text{LSM}_{120}}$ coincide because we use a square box, $L_x=L_y$. This degeneracy is slightly lifted by qubit paddles. Nevertheless, to  calculate $g_{210}$ numerically, we slightly increased $L_x$ to 30.2 mm to break the degeneracy. 

The ratio $A_{210}/A$ between the maximum couplings is 3.19 numerically and 3.23 using the LSM-mode analytics, further confirming our physics understanding. Both ratios are close to $2\omega_{210}/\omega_{110}=3.15$, the formula which follows from Eq.\ (\ref{eq:Ex_LSM-t}) with constant $E_{0,x}$. 

Finally, let us briefly discuss why the spatial profile of couplings, $\propto \cos (a\pi x/L_x)\sin (b\pi y/L_y)$, is proportional to the $x$-derivative of the TM mode, $\partial_x E_z^{\text{TM}_{ab0}}$. As follows from the discussion in Appendix \ref{appendix:LSM_modes_derivation}, the $E_x$ component of the LSM electric field is proportional to the second derivative of  $E_z$ field, $\partial_x\partial_z E_z^{\text{LSM}_{ab0}}$. However, $E_z^{\text{LSM}_{ab0}}$ and $E_z^{\text{TM}_{ab0}}$ have the same $x$ and $y$ dependence; the main difference is that $E_z^{\text{LSM}_{ab0}}$ has a weak $z$-dependence and the jump at the dielectric surface. As a result, $\partial_x\partial_z E_z^{\text{LSM}_{ab0}}$ has the same $x$ and $y$ dependence as $\partial_x E_z^{\text{TM}_{ab0}}$, thus explaining why the $g$-coupling profile is proportional to $\partial_x E_z^{\text{TM}_{ab0}}$.

\section{Conclusion}
In this work, we have compared four full-wave electromagnetic methods for extracting the coupling $g$ between localized qubit modes and distributed package modes: avoided crossing, energy participation ratio, induced EMF, and impedance matrix. This suite of methods provides a flexible computational framework, offering specific techniques tailored for either eigenmode or driven full-wave electromagnetic solvers. 

We have applied these techniques to a $10\times10$ transmon array in a sample processor and demonstrated a quantitative agreement, with a maximum relative deviation of less than $5\%$ across all four techniques. We have also shown that the spatial profile of these couplings across the processor can be understood via  the interaction between the transmon's electric dipole moment and the in-plane electric field component of the package modes.

Understanding the spatial dependence of the couplings between qubits and package modes provides a systematic basis for engineering geometric coupling nulls and evaluating crosstalk, informing the design and integration of dense superconducting quantum processors.

\appendix

\section{Hamiltonian, eigenfrequencies, and avoided level crossing}
\label{app:avoided_crossing}

In this Appendix, we consider the linear two-resonator system shown in Fig.\ \ref{fig:summary_of_methods}(b) and its generalization to several coupled single-ended (grounded) resonators. For differential qubits, we consider the effective single-ended circuits; qubit nonlinearity is neglected. 
We discuss both the quantum Hamiltonian approach and the classical eigenfrequency approach, leading to the exact and approximate results for the avoided crossing.

\subsection{Quantum treatment}

For a system of coupled single-ended linear LC resonators, 
the classical capacitive and inductive energies are 
    \begin{eqnarray}
&& \mathcal{E}_{\rm cap} = \frac{1}{2}V^T\mathbf{C} V = \frac{1}{2} Q^T \mathbf{C}^{-1} Q,  
\\ 
&& \mathcal{E}_{\rm ind}=\frac{1}{2} I^T \mathbf{L} I = \frac{1}{2} \Phi^T  \mathbf{L}^{-1} \Phi,  
    \end{eqnarray}
where $V$, $Q$, $I=-\dot{Q}$, and $\Phi=\int V(t) \, dt$ are the column vectors of nodal voltages, nodal charges, inductor currents, and inductor magnetic fluxes, respectively, the superscript $T$ means transpose, while $\mathbf{C}$ and  $\mathbf{L}$ are the symmetric capacitance and inductance matrices defined via the standard relations $Q=\mathbf{C} V$ and $\Phi = \mathbf{L} I$. In particular, for the circuit in Fig.\ \ref{fig:summary_of_methods}(b), 
    \begin{equation}
\mathbf{C} =\begin{pmatrix}
    C_m+C_c & -C_c
    \\
    -C_c & C_n+C_c 
\end{pmatrix}  , \quad 
\mathbf{L} =\begin{pmatrix}
    L_m & M
    \\
    M & L_n 
\end{pmatrix} , 
    \label{eq:C-L-matrices}\end{equation}
while $Q^T=(Q_m, Q_n )$ and $\Phi^T=(\Phi_m, \Phi_n)$.

Choosing the fluxes $\Phi$ as generalized coordinates, we can write the Lagrangian as 
\begin{equation}
    \mathcal{L} = \frac{1}{2}\dot{\Phi}^T \mathbf{C} \dot{\Phi} - \frac{1}{2}\Phi^T \mathbf{L}^{-1}\Phi,
\label{eq:Lagrangian-class}\end{equation} 
since $\dot{\Phi}=V$. 
The conjugate momenta in this case are the nodal charges $Q$, because $\delta\mathcal{L}/\delta\dot{\Phi} = \mathbf{C}\dot{\Phi}=Q$. To find the Hamiltonian $H$, we use the Legendre transformation $H = Q^T \dot{\Phi} - \mathcal{L}$, thus obtaining the total energy in terms of coordinates $\Phi$ and conjugate momenta $Q$, 
\begin{equation}
    H = \frac{1}{2} Q^T \mathbf{C}^{-1}Q + \frac{1}{2} \Phi^T \mathbf{L}^{-1}\Phi.
\label{eq:Ham-class}\end{equation}
In this Hamiltonian, the diagonal terms in matrices $\mathbf{C}^{-1}$ and $\mathbf{L}^{-1}$ describe the ``bare'' resonators, while the off-diagonal terms describe their charge-charge (capacitive) and flux-flux (inductive) couplings. From the diagonal terms, it is easy to see that the bare resonators have ``masses'' $m_k=[(\mathbf{C}^{-1})_{kk}]^{-1}$ and frequencies 
    \begin{equation}
        \omega_k = \sqrt{(\mathbf{L}^{-1})_{kk}(\mathbf{C}^{-1})_{kk}} \, .
    \label{eq:bare-freq}\end{equation}

\vspace{0.2cm}

For the quantum treatment of this system, we follow the standard circuit
quantization: \cite{Devoret1997} we use the same Hamiltonian
(\ref{eq:Ham-class}) and promote the coordinates $\Phi$ and conjugate
momenta $Q$ to operators, $\Phi\rightarrow\hat{\Phi}$ and
$Q\rightarrow\hat{Q}$, satisfying
$[\hat{Q}_k,\hat{\Phi}_l]=-i\hbar\delta_{kl}$. As the next step, for convenience, we express  $\hat{\Phi}$ and $\hat{Q}$ via the dimensionless creation and annihilation operators introduced for {\it bare} resonators:  
\begin{equation}
    \hat{\Phi}_k = \Phi_k^{\text{zpf}}(\hat{a}_k + \hat{a}_k^\dagger), 
    \quad 
    \hat{Q}_k = -i Q_k^{\text{zpf}}(\hat{a}_k - \hat{a}_k^\dagger),
\label{eq:Phi-Phi-zpk}\end{equation}
where $[\hat{a}_k, \hat{a}^\dagger_l]=\delta_{kl}$ and the dimensional coefficients $\Phi_k^{\text{zpf}}$ and $Q_k^{\text{zpf}}$ are the zero-point fluctuations, i.e., 
the ground state width (standard deviation of the probability distribution) for coordinate $\Phi_k$ and momentum $Q_k$ of $k$th bare resonator. From elementary quantum mechanics, $\Phi_k^{\text{zpf}}=\sqrt{\hbar/2m_k\omega_k}$ and $Q_k^{\text{zpf}}=\sqrt{\hbar m_k\omega_k/2}$, 
which can also be expressed via the impedance $Z_k$,\cite{Devoret1997} 
\begin{equation}
 \Phi_k^{\text{zpf}} = \sqrt{\frac{\hbar Z_k}{2}}\, , \,\,\, Q_k^{\text{zpf}} = \sqrt{\frac{\hbar}{2Z_k}}\, , \,\,\, Z_k = \sqrt{\frac{(\mathbf{C}^{-1})_{kk}}{ (\mathbf{L}^{-1})_{kk}}} \, .   
\label{eq:Phi-k-zpf}\end{equation}

Rewriting the Hamiltonian (\ref{eq:Ham-class}) with substitution  (\ref{eq:Phi-Phi-zpk}), using Eqs.\ (\ref{eq:bare-freq}) and (\ref{eq:Phi-k-zpf}), and subtracting the unimportant constant $\sum_k \hbar\omega_k/2$, we obtain 
\begin{align}
    & \frac{\hat{H}}{\hbar} = \sum\nolimits_k \omega_k\hat{a}_k^\dagger\hat{a}_k 
    - \sum\nolimits_{k<l} g^C_{kl}(\hat{a}_k - \hat{a}_k^\dagger)(\hat{a}_l - \hat{a}_l^\dagger) \nonumber \\ 
    & \hspace{0.7cm} - \sum\nolimits_{k<l} g^L_{kl}(\hat{a}_k + \hat{a}_k^\dagger)(\hat{a}_l + \hat{a}_l^\dagger),
    \label{eq:Ham-quant}\\ 
    & g^C_{kl} = \frac{ k^C_{kl}}{2} \sqrt{\omega_k\omega_l}, \,\,\,  k^C_{kl} = \frac{(\mathbf{C}^{-1})_{kl}}{\sqrt{(\mathbf{C}^{-1})_{kk}(\mathbf{C}^{-1})_{ll}}} \, , \quad 
    \label{eq:g-C-kl}\\
       & g^L_{kl} = \frac{k^L_{kl}}{2}  \sqrt{\omega_k\omega_l}, \,\,\,  k^L_{kl} = -\frac{(\mathbf{L}^{-1})_{kl}}{\sqrt{(\mathbf{L}^{-1})_{kk}(\mathbf{L}^{-1})_{ll}}} \, , \quad 
 \label{eq:g-L-kl}\end{align}
where the bare resonator frequencies $\omega_k$ are given by Eq.\ (\ref{eq:bare-freq}) and the coefficients $k^C_{kl}$ and $k^L_{kl}$ can be called the capacitive and inductive coupling efficiencies between $k$th and $l$th resonators. Note that the minus sign for the capacitive-coupling term in Eq.\ (\ref{eq:Ham-quant}) comes from the imaginary unit in Eq.\ (\ref{eq:Phi-Phi-zpk}), so that the resulting excitation-preserving terms ($\hat{a}_k^\dagger  \hat{a}_l +\hat{a}_k  \hat{a}^\dagger_l$) have positive coefficient $g^C_{kl}$. In contrast, the minus sign for the inductive-coupling term in Eq.\ (\ref{eq:Ham-quant}) is introduced for convenience and is compensated by the negative sign for $k^L_{kl}$ in Eq.\ (\ref{eq:g-L-kl}), so that $k^L_{kl}$ is positive in the typical case when $\mathbf{L}_{kl}$ is positive and $(\mathbf{L}^{-1})_{kl}$ is negative. 

\vspace{0.2cm}

For the two-resonator case shown in Fig.\ \ref{fig:summary_of_methods}(b), the capacitance and inductance matrices are given by Eq.\ (\ref{eq:C-L-matrices}),  so the Hamiltonian (\ref{eq:Ham-quant})--(\ref{eq:g-L-kl}) reduces to Eq.\ (\ref{eq:hamiltonian}):  
\begin{align}
&   \frac{\hat{H}}{\hbar} = \omega_m\hat{a}_m^\dagger\hat{a}_m + \omega_n\hat{a}_n^\dagger\hat{a}_n 
    - g_C(\hat{a}_m - \hat{a}_m^\dagger)(\hat{a}_n - \hat{a}_n^\dagger) 
\nonumber \\
    &\hspace{0.7cm} - g_L(\hat{a}_m + \hat{a}_m^\dagger)(\hat{a}_n + \hat{a}_n^\dagger),
 \label{eq:hamiltonian-2} \\ 
      & g_{C} = \frac{k_C}{2} \sqrt{\omega_m\omega_n}\, , \quad  g_L = \frac{k_{L}}{2}  \sqrt{\omega_m\omega_n} \, , 
    \label{eq:g-C-g-L}\\
&    k_{C} = \frac{C_c}{\sqrt{(C_m +C_c)(C_n+C_c)}} \, ,  \,\,\, k_L = \frac{M}{\sqrt{L_m L_n}} \, , 
    \label{eq:k-C-k-L}\\
&  \omega_m = \Big[ L_m \Big( 1-\frac{M^2}{L_m L_n}\Big) \Big( C_m+\frac{C_c C_n}{C_c+C_n}\Big)\Big]^{-1/2} , 
\label{eq:omega-m}\\
& \omega_n = \Big[ L_n \Big( 1-\frac{M^2}{L_m L_n}\Big) \Big( C_n+\frac{C_c C_m}{C_c+C_m}\Big)\Big]^{-1/2} . 
\label{eq:omega-n}\end{align}
Note that the mutual inductance $M$ is typically small, so the quadratic correction in $M$ can usually be neglected, $1-M^2/L_m L_n \approx 1$. Similarly, neglecting the second-order correction in the small coupling capacitance $C_c$, we can approximate the bare frequencies as $\omega_m\approx 1\sqrt{L_m (C_m+C_c)}$ and $\omega_n\approx 1\sqrt{L_n (C_n+C_c)}$. 
% (For the multi-resonator case, the similar approximation would replace Eq.\ (\ref{eq:bare-freq}) with $\omega_k\approx 1/\sqrt{\mathbf{L}_{kk}\mathbf{C}_{kk}}$.) 

\vspace{0.2cm}

The exact eigenfrequencies (eigenvalues) of the Hamiltonian (\ref{eq:hamiltonian-2}) [or its generalization (\ref{eq:Ham-quant})] coincide with the classical eigenfrequencies, discussed in the next section. It is possible to find eigenfrequencies of Eq.\ (\ref{eq:hamiltonian-2}) directly (see e.g., Ref.\
\onlinecite{Xiao2009}), but the analysis greatly simplifies when $|\omega_m-\omega_n|\ll \omega_m+\omega_n$ and $|g_{C}|+|g_L|\ll \omega_{m}+\omega_n$. In this case we can use the rotating-wave approximation (RWA) and neglect the terms $\hat{a}_m\hat{a}_n$ and $\hat{a}_m^\dagger\hat{a}_n^\dagger$, which do not preserve the number of excitations. Then the Hamiltonian (\ref{eq:hamiltonian-2}) reduces to 
\begin{equation}
    \frac{\hat{H}}{\hbar} \approx (\hat{a}_m^\dagger \,\,\, \hat{a}_n^\dagger)
    \begin{pmatrix}
        \omega_m & g_C - g_L  \\
        g_C - g_L \,\,\,\,  & \omega_n
    \end{pmatrix}
    \begin{pmatrix}
        \hat{a}_m \\
        \hat{a}_n
    \end{pmatrix},  
    \label{eq:2-level-Ham}\end{equation}
which can be easily diagonalized to obtain the eigenfrequencies 
\begin{equation}
    \omega_\pm = \frac{\omega_m+\omega_n}{2} \pm \sqrt{\frac{(\omega_n-\omega_m)^2}{4} + (g_C - g_L)^2} \, .
    \label{eq:app_quantum_omega_pm}
\end{equation}
This is the standard avoided level crossing behavior, in which the capacitive and inductive couplings come in the combination $g = g_C - g_L$. The minimum separation of the eigenfrequencies $\omega_+$ and $\omega_-$ occurs at $\omega_m = \omega_n$, when 
\begin{equation}
  \min (\omega_+ - \omega_- ) = 2\, |g_C-g_L| \, .
    \label{eq:app_quantum_separation}
\end{equation}

We emphasize that Eqs.\ (\ref{eq:2-level-Ham})--(\ref{eq:app_quantum_separation}) are approximate (using RWA), while Eqs.\ (\ref{eq:Ham-quant})--(\ref{eq:omega-n}) are exact. The exact eigenfrequencies in the quantum approach coincide with the classical eigenfrequencies, derived in the next section. Let us briefly discuss how to obtain them in the quantum approach. Introducing the new coordinates $X={\mathbf C}^{1/2}\Phi$, we can write the Lagrangian (\ref{eq:Lagrangian-class}) as     $\mathcal{L} = \frac{1}{2}\dot{X}^T \dot{X} - \frac{1}{2}X^T \mathbf{C}^{-1/2}\mathbf{L}^{-1}\mathbf{C}^{-1/2} X$, from which the conjugate momentum is $P=\dot{X}$. Therefore, the corresponding quantum Hamiltonian is $\hat{H}=\frac{1}{2}\hat{P}^T \hat{P} + \frac{1}{2}\hat{X}^T \mathbf{C}^{-1/2}\mathbf{L}^{-1}\mathbf{C}^{-1/2} \hat{X}$ with the commutation relation $[\hat{P}_{k},\hat{X}_l]=-i\hbar\delta_{kl}$. Further introducing the ``rotated'' coordinates $\hat{\tilde{X}}$ and conjugate momenta $\hat{\tilde{P}}$, defined by $\hat{X}=\mathbf{U}\hat{\tilde{X}}$ and  $\hat{P}=\mathbf{U}\hat{\tilde{P}}$, where $\mathbf{U}$ is a real-valued unitary (orthogonal) matrix, $\mathbf{U}^T=\mathbf{U}^\dagger=\mathbf{U}^{-1}$, we get the same commutation relations, $[\hat{\tilde P}_{k},\hat{\tilde X}_l]=-i\hbar\delta_{kl}$ and the Hamiltonian 
$\hat{H}=\frac{1}{2}\hat{\tilde P}^T \hat{\tilde P} + \frac{1}{2}\hat{\tilde X}^T \mathbf{U}^{-1} (\mathbf{C}^{-1/2}\mathbf{L}^{-1}\mathbf{C}^{-1/2}) \mathbf{U} \hat{\tilde X}$.
Therefore, by diagonalizing the real-valued symmetric matrix $\mathbf{C}^{-1/2}\mathbf{L}^{-1}\mathbf{C}^{-1/2}$ (so that ${\mathbf U}$ is the corresponding column-wise eigenvector matrix), we can obtain a fully diagonal Hamiltonian, from which we immediately obtain the eigenfrequencies. Thus we obtain the result for exact  eigenfrequencies in the quantum approach: the squared eigenfrequencies are equal to the eigenvalues of the matrix $\mathbf{C}^{-1/2}\mathbf{L}^{-1}\mathbf{C}^{-1/2}$.

\subsection{Classical treatment}

For the system of several single-ended resonators considered in the previous section, we can write the following evolution equations: 
    \begin{equation}
        \ddot{Q}=-\dot{I}=-\mathbf{L}^{-1}\dot\Phi = -\mathbf{L}^{-1}V = - \mathbf{L}^{-1} \mathbf{C}^{-1}Q.   
    \end{equation}
Therefore, the eigenfrequencies $\omega$ can be obtained from the equation 
    \begin{equation}
      (\mathbf{L}^{-1} \mathbf{C}^{-1}) \, Q =\omega^2 Q.   
    \end{equation}
We can similarly derive the equation $\ddot{V}=- \mathbf{C}^{-1} \mathbf{L}^{-1} V$, from which the eigenfrequencies  $\omega$  can be found from 
    \begin{equation}
      (\mathbf{C}^{-1}\mathbf{L}^{-1}) \, V =\omega^2 V.   
    \end{equation} 
    
Note that the eigenvalues of the matrices $\mathbf{L}^{-1} \mathbf{C}^{-1}$ and $\mathbf{C}^{-1}\mathbf{L}^{-1}$ coincide; this is a general property of products of any square matrices. Because of this property, the matrix $\mathbf{C}^{-1/2}\mathbf{L}^{-1}\mathbf{C}^{-1/2}$ obtained in the quantum derivation above, also has the same eigenvalues, thus proving that the classical and quantum approaches give the same eigenfrequencies. 

The equivalent equations for the eigenfrequencies are 
   \begin{equation}
      (\mathbf{C} \mathbf{L}) \, Q =\omega^{-2} Q, \,\,\,   
       (\mathbf{L} \mathbf{C}) \, V =\omega^{-2} V. 
    \label{eq:omega-2-eigen}\end{equation}

\vspace{0.2cm}

In particular, for the two-resonator case shown in Fig.\ \ref{fig:summary_of_methods}(b), the matrices $\mathbf{C}$ and $\mathbf{L}$ are given in Eq.\ (\ref{eq:C-L-matrices}). Using the first equation in (\ref{eq:omega-2-eigen}), we can find the eigenfrequencies $\omega_\pm$ from the condition $\det (\mathbf{C} \mathbf{L} -\omega_\pm^{-2} \mathbf{I})=0$, where $\mathbf{I}$ is the identity matrix. This gives us the equation 
\begin{align}
  &  \omega_\pm^4 (L_m L_n-M^2)[C_nC_m+C_c(C_m+C_n)] 
  \nonumber \\
 &- \omega_\pm^2  [L_m(C_m+C_c)  +L_n(C_n+C_c)-2MC_c]+1=0, \quad\quad
\label{eq:omega-pm-LC}\end{align}
from which we can easily find the eigenfrequencies in terms of capacitances and inductances. However, it is more instructive to express $\omega_\pm$ in terms of the bare frequencies $\omega_m$, $\omega_n$ and coupling efficiencies $k_C$, $k_L$, given by Eqs.\ (\ref{eq:k-C-k-L})--(\ref{eq:omega-n}). For that we rewrite Eq.\ (\ref{eq:omega-pm-LC}) as 
\begin{align}
&\left(\frac{\omega_\pm^2}{\omega_m\omega_n}\right)^2 - \frac{\omega_\pm^2}{\omega_m\omega_n} \left(\frac{\omega_m^2+\omega_n^2}{\omega_m\omega_n} - 2k_Ck_L\right) 
    \nonumber\\
& \hspace{0.0cm} +(1-k_C^2)(1-k_L^2)=0, 
\end{align}
which gives us the eigenfrequencies $\omega_\pm$:
\begin{align}
    & \frac{\omega_\pm^2}{\omega_m\omega_n} = \frac{\omega_m^2 +\omega_n^2}{2\omega_m\omega_n} -k_C k_L \pm \bigg[ \left( \frac{\omega_m^2-\omega_n^2}
    {2\omega_m\omega_n}\right)^2 
    \nonumber \\
    & \hspace{1.4cm} +k_C^2 +k_L^2   - \frac{\omega_m^2+\omega_n^2}{\omega_m\omega_n} \, k_C k_L  \bigg]^{1/2} . 
\label{eq:omega-pm-kc-kl}\end{align}

Comparing this exact result with the quantum avoided crossing result (\ref{eq:app_quantum_omega_pm}), we see that Eq.\ (\ref{eq:omega-pm-kc-kl}) indeed reduces to Eq.\ (\ref{eq:app_quantum_omega_pm}) if $\omega_m/\omega_n \approx 1$, $|k_C| \ll 1$, and $|k_L|\ll 1$. The difference between the two results is because of the Rotating Wave Approximation used to derive Eq.\ 
(\ref{eq:app_quantum_omega_pm}).

In the case of exact resonance, $\omega_m=\omega_n$, we get
\begin{equation}
    \omega_\pm =\omega_m \sqrt{1 -k_C k_L \pm |k_C-k_L|} \, , 
\end{equation}
which for $|k_C|\ll 1$ and $|k_L|\ll 1$ reduces to the quantum result (\ref{eq:app_quantum_separation}),
    \begin{equation}
        \omega_+ -\omega_- \approx 2 |g_C-g_L|. 
    \end{equation}

\section{Energy-participation-ratio derivations}
\label{app:epr}

In this Appendix, we derive the relation in Eq.~(\ref{eq:g_tot}) between the energy participation ratio and the qubit-cavity couplings.

Consider the equivalent circuit in Fig. \ref{fig:summary_of_methods}(b), comprising two LC resonators coupled by a mutual inductance $M$ and a coupling capacitor $C_c$; resonator $n$ represents qubit mode and resonator $m$ represents cavity/package mode. Following the procedure of the EPR method, we consider that cavity eigenmode is excited at its eigenfrequency, which is approximately equal to bare mode frequency $\omega_m$.  Using the node flux $\Phi$ (where $V = j\omega_m\Phi$) as the variable and applying Kirchhoff's Current Law (KCL) at node $n$ we have:
\begin{equation}
    -\omega_m^2 C_n \Phi_n - \omega^2 C_c (\Phi_n - \Phi_m) + I_{L_n} = 0
\end{equation}
Assuming weak inductive coupling ($M \ll \sqrt{L_m L_n}$), the current through the $n$-th inductor is $I_{L_n} \approx \Phi_n/L_n - [M/(L_m L_n)]\Phi_m$. Substituting this into KCL, separating the terms, and using the approximation of bare qubit frequency $\omega_n$ introduced in Appendix~\ref{app:avoided_crossing}, i.e. $\omega_n\approx 1\sqrt{L_n (C_n+C_c)}$, we solve for the node flux ratio:
\begin{equation}
\label{eq:node_flux_ratio}
    \frac{\Phi_n}{\Phi_m} = \frac{M/(L_m L_n) - \omega_m^2 C_c}{(C_{n}+C_{c})(\omega_n^2 - \omega_m^2)} \, . 
\end{equation}
By definition, the energy participation ratio $p_{mn}$ for a transmon evaluates the inductive energy of the Josephson junction, not the total node energy [see Eq.~(\ref{eq:pmn})]. Because the junction resides in an inductive loop, the mutual inductance induces a flux shift. Neglecting geometric self-inductance of qubit (typically much smaller than junction inductance), the true flux across the junction is $\Phi_{\rm n,\rm J} = L_n I_{L_n} = \Phi_n - (M/L_m) \Phi_m$. Substituting this relation into Eq.~(\ref{eq:node_flux_ratio}) yields the ratio between junction flux $\Phi_{\rm n,\rm J}$ and cavity node flux $\Phi_m$:~\cite{minev2021}
\begin{align}
    \frac{\Phi_{\rm n,\rm J}}{\Phi_m} & =  \frac{M/(L_m L_n) - \omega_m^2 C_c}{(C_{n}+C_{c})(\omega_n^2 - \omega_m^2)} - \frac{M}{L_m}
    \nonumber \\
    & = \frac{\omega_m^2 \left[ M/L_m - C_c/(C_{n}+C_{c}) \right]}{\omega_n^2 - \omega_m^2} \, . 
    \label{eq:flux_ratio}
\end{align}
To relate the ratio $\Phi_{\rm n, \rm J}/\Phi_m$ to $p_{mn}$, we use its definition in Eq.~(\ref{eq:pmn}) and find the following: 
\begin{align}
    p_{mn} \equiv \frac{\mathcal{E}_{\rm n,\rm JJ}}{\mathcal{E}_{m,\rm ind}}\approx \frac{\Phi_{\rm n, \rm J}^2/L_n}{\Phi_m^2/L_m},
\label{eq:pmn_approx}
\end{align}
where in the second step we have approximated the total inductive energy in the mode $m$ to be $\mathcal{E}_{m,\rm ind}\approx \Phi_m^2/2L_m$.
%As explained in Sec.~\ref{sec:EPR_method}, the Energy Participation Ratio $p_{mn}$ is defined as the ratio of the inductive energy stored in junction $n$ and the total inductive energy of mode $m$. To the lowest-order in the qubit-mode coupling, it is given by $p_{mn} = (L_m/L_n)(\Phi_{J,n}/\Phi_m)^2$. 

Finally, substituting Eq.~(\ref{eq:flux_ratio}) into Eq.~(\ref{eq:pmn_approx}), and using Eqs.~(\ref{eq:gs}) and (\ref{eq:ks}) to express the coupling inductance and capacitance in terms of couplings $g_C,g_L$, we arrive at Eq.~(\ref{eq:g_tot}) in the main text (repeated here):
\begin{equation}
    \left| g_C - g_L \frac{\omega_m}{\omega_n} \right| = \frac{|\omega_m^2 - \omega_n^2|}{2\sqrt{\omega_m \omega_n}} \sqrt{p_{mn}}.
\end{equation}

\section{Induced-EMF method derivations}
\label{app:induced_voltage}
In this Appendix, we derive the relation in Eq.~(\ref{eq:v_o}) between qubit's voltage $V_{n,\rm open}$ across opened junction and qubit-cavity coupling $g$. 

Let us consider the circuit in Fig.~\ref{fig:schematics_detailed}. An opened qubit junction corresponds to the limit of $L_n'\rightarrow \infty$. To derive Eq.~(\ref{eq:v_o}), similar to the EPR derivation in Appendix~\ref{app:epr}, we resonantly excite the cavity mode $m$ with total energy $\mathcal{E}_m$ such that the cavity node has a voltage $V_m$. The cavity node refers to the node that connects cavity capacitance $C_m$ and coupling capacitance $C_c$. By Kirchhoff's circuit laws, the voltage across qubit port $V_{n,\rm open}$ can be expressed in terms of $V_m$ through the following relation:
\begin{equation}
    V_{n,\rm open} = C_cV_m/(C_{n}+C_{c}) - MV_m/L_m, \label{eq:V_open}
\end{equation}
where $L_m = L_m' + \delta L_m'$. 
The first term on the right-hand side of Eq.~(\ref{eq:V_open}) is the voltage across the qubit capacitance $C_n$ derived through voltage division of $V_m$ between $C_c$ and $C_n$. The second term is the voltage across qubit geometric inductance $\delta L_n'$ induced by the inductive coupling $M$. We subtract the second term from the first term to obtain the voltage across the qubit port. 

Using Eqs.~(\ref{eq:gs}) and (\ref{eq:ks}), we can relate the right-hand side of Eq.~(\ref{eq:V_open}) to the capacitive and inductive coupling when qubit is tuned on resonance with cavity mode:
\begin{equation}
    V_{n,\rm open} = \frac{2V_m}{\omega_m}\sqrt{\frac{C_{m}+C_{c}}{C_{n}+C_{c}}}\left(g_C|_{\omega_n\rightarrow \omega_m} - g_L|_{\omega_n\rightarrow \omega_m}\right).
\label{eq:g_to_v_o}
\end{equation}
Note that by definition [see Eq.~(\ref{eq:g-total})], the sought-after total coupling $g$ is equal to $g=g_C|_{\omega_n\rightarrow \omega_m} - g_L|_{\omega_n\rightarrow \omega_m}$. Finally, we express cavity voltage $V_m$ in Eq.~(\ref{eq:g_to_v_o}) in terms of the total energy $\mathcal{E}_m$ via the relation $\mathcal{E}_m = 1/2\, (C_{m}+C_{c}) |V_m|^2$. This leads to Eq.~(\ref{eq:v_o}) in the main text (repeated here):
\begin{equation}
    g  = \frac{\omega_m V_{n,\, \rm open}}{2} \sqrt{\frac{C_{n}+C_{c} }{2\mathcal{E}_m}}.
\end{equation}

Similarly, it can be shown that if we replace the qubit's junction with a short-circuit boundary condition, the current through the junction, $I_{n,\rm short}$, is related to the on-resonance couplings as the following:
\begin{equation}
    g_C|_{\omega_n\rightarrow \omega_m} - g_L|_{\omega_n\rightarrow \omega_m}  = \frac{I_{n,\rm short}}{2\sqrt{2}} \sqrt{\frac{1}{ (C_{n}+C_{c}) \mathcal{E}_m}}.
\end{equation}
The above relation provides an alternative way of extracting coupling between the qubit and cavity mode.

\section{Impedance matrix method fitting procedure and derivations}
\label{app:lc_circuit}
\subsection{Fitting procedure}
In this section, we explain the procedure for obtaining the matrices $\mathbf{C}$, $\mathbf{L}'$ and $\delta \mathbf{L}'$ in Eq.~(\ref{eq:z_fit_new}) via fitting to the simulated impedance matrix.

We first rewrite Eq.~(\ref{eq:z_fit_new}) in a form that is more convenient for fitting: 
\begin{align}
\label{eq:Z_convenient}
\mathbf{Z}\left(\omega\right)&=\mathbf{Z}^{(0)}\left(\omega\right)+j\omega\widetilde{\delta\mathbf{L}}, \\
\mathbf{Z}^{(0)}\left(\omega\right) &= \left[(j\omega)^{-1}\mathbf{\widetilde L}^{-1}+j\omega\mathbf{\widetilde C}\right]^{-1},
\label{eq:z_pole}
\end{align}
where the matrices $\mathbf{\widetilde L}, \mathbf{\widetilde C}$ and $\widetilde{\delta\mathbf{L}}$ are related to $\mathbf{C}, \mathbf{L}'$ and $\delta \mathbf{L}'$ via the following relations:
\begin{align}
\mathbf{C}=~&\boldsymbol{\alpha}^{-1}\mathbf{\widetilde C}(\boldsymbol{\alpha}^{-1})^{T},\quad \mathbf{L}^{\prime-1}=\boldsymbol{\alpha}^{-1}\mathbf{\widetilde L}^{-1},\nonumber \\\delta\mathbf{L}^{\prime}=~&\boldsymbol{\alpha}^{T}\widetilde{\delta\mathbf{L}},\quad \boldsymbol{\alpha}\equiv\mathbf{I}+\mathbf{\widetilde L}^{-1}\widetilde{\delta\mathbf{L}}.
\label{eq:lc_to_lc_prime}
\end{align}

As we will explain in next section, the expression for the impedance matrix in Eq.~(\ref{eq:Z_convenient}) has a clear physical interpretation: the first term $\mathbf{Z}^{(0)}$ is simply the impedance matrix of a network of inductors and capacitors; the second term can be understood as a correction to the first term to capture the tails from high-frequency modes that are not captured by $\mathbf{Z}^{(0)}$ due to its finite frequency range and finite dimension.

Impedance matrix $\mathbf{Z}^{(0)}(\omega)$ in Eq.~(\ref{eq:z_pole}) possesses a set of poles located at eigenfrequencies determined by the capacitance and inductance matrices $\mathbf{\widetilde C}$ and $\mathbf{\widetilde L}$ [see Eq.~(\ref{eq:eigenvalue_equation})].
For an $N \times N$ matrix, $\mathbf{Z}^{(0)}(\omega)$ generally has $N$ poles. Here, we focus on the case of $N = 2$, where one pole corresponds to the qubit mode, and the other corresponds to the package mode. Locations and residues at the poles are determined by the capacitance and inductance matrices $\mathbf{\widetilde C}$ and $\mathbf{\widetilde L}$. Analyzing the simulated impedance matrix near the poles allows us to extract these matrices. 

In the following, we give the mathematical results that are needed to perform this extraction, and leave the detailed derivation to the next section. First, one can show that $\mathbf{Z}^{(0)}$ can be written as the following summation:
\begin{align}
\label{eq:sum_over_poles}
\mathbf{Z}^{(0)}\left(\omega\right)=\sum_{\mu=1}^N\frac{2j\omega\omega_{\mu}}{\omega^{2}_{\mu}-\omega^{2}}\mathbf{R}_{\mu},
\end{align}
where mode frequency $\omega_{\mu}$ sets the location of the pole, and $\mathbf{R}_{\mu}$ is the residue matrix at the pole. We fit Eq.~(\ref{eq:sum_over_poles}) to the simulated impedance matrix near the poles to get both $\omega_{\mu}$ and $\mathbf{R}_{\mu}$ for each $\mu$.

Next, we compute $\mathbf{\widetilde C}$ and $\mathbf{\widetilde L}$ from the fitted $\omega_{\mu}$ and $\mathbf{R}_{\mu}$.
One can theoretically show that residue matrix can be written as a product of a column vector (which we call $\boldsymbol{\sigma}_{\mu}$) and its transpose: 
\begin{align}
\label{eq:residue_matrix}
    \mathbf{R}_{\mu}=\boldsymbol{\sigma}_{\mu}\boldsymbol{\sigma}^{T}_{\mu}.
\end{align}
By stacking all the column vectors $\boldsymbol{\sigma}_{\mu}$, we build an $N \times N$ matrix
\begin{align}
\label{eq:sigma_matrix_stack}
    \boldsymbol{\Sigma} = (\boldsymbol{\sigma}_{1}, \boldsymbol{\sigma}_{2}, \ldots,\boldsymbol{\sigma}_{N}).
\end{align}
One can show that the matrix $\boldsymbol{\Sigma}$ relates to $\mathbf{\widetilde C}$ and $\mathbf{\widetilde L}$ via the following relations:
\begin{equation}
\label{eq:sigma_to_lc}
    \mathbf{\widetilde L}^{-1} = \frac{1}{2} \, \boldsymbol{\Sigma}^{-T} \hat{\bm{\Omega}} \boldsymbol{\Sigma}^{-1}, \quad \mathbf{\widetilde C} = \frac{1}{2} \, \boldsymbol{\Sigma}^{-T} \hat{\bm{\Omega}}^{-1} \boldsymbol{\Sigma}^{-1} \,,
\end{equation}
where $\hat{\boldsymbol{\Omega}}$ is a diagonal matrix containing the mode frequencies $\omega_{\mu}$:
\begin{equation}
\label{eq:Omega}
    \hat{\boldsymbol{\Omega}} = \begin{pmatrix}
  \omega_1 &  & &  \\
   & \omega_2 &  &  \\
   &  & \ddots &  \\
   &  &  & \omega_N
\end{pmatrix} \,.
\end{equation}
We use Eqs.~(\ref{eq:residue_matrix},\ref{eq:sigma_matrix_stack},\ref{eq:sigma_to_lc},\ref{eq:Omega}) to get $\mathbf{\widetilde C}$ and $\mathbf{\widetilde L}$ from  $\omega_{\mu}$ and $\mathbf{R}_{\mu}$.

Having explained how to obtain $\mathbf{\widetilde C}$ and $\mathbf{\widetilde L}$ in Eq.~(\ref{eq:Z_convenient}), we now explain how to obtain $\widetilde{\delta\mathbf{L}}$. The idea is to subtract $\mathbf{Z}^{(0)}$ from the simulated impedance matrix and fit this difference to a linear function $j\omega\widetilde{\delta\mathbf{L}}$. The fitted slope immediately gives us $\widetilde{\delta\mathbf{L}}$. Having extracted $\mathbf{\widetilde C}$, $\mathbf{\widetilde L}$ and $\widetilde{\delta\mathbf{L}}$, we use Eq.~(\ref{eq:lc_to_lc_prime}) to obtain the sought-after $\mathbf{C}$, $\mathbf{L}'$ and $\delta \mathbf{L}'$.

\subsection{Derivations}
In this section, we first give the derivations for Eqs.~(\ref{eq:sum_over_poles},\ref{eq:residue_matrix},\ref{eq:sigma_to_lc}), and then elaborate on the origin of the $\widetilde{\delta \mathbf{L}}$ term in Eq.~(\ref{eq:Z_convenient}).

The poles of the impedance matrix $\mathbf{Z}^{(0)}(\omega)$ in Eq.~(\ref{eq:z_pole}) are described by an eigenvalue problem:
\begin{equation}
    \mathbf{\widetilde C}^{-1}\mathbf{\widetilde L}^{-1}\mathbf{v}_{\mu} = \omega^2_{\mu}\mathbf{v}_{\mu},
\label{eq:eigenvalue_equation}
\end{equation}
featuring eigenmode frequencies $\omega_{\mu}$ and voltage eigenvectors $\mathbf{v}_{\mu}$ for each mode $\mu$. 

Applying a coordinate transformation $\mathbf{v}_{\mu} = \mathbf{\widetilde C}^{-1/2}\mathbf{u}_{\mu}$ to Eq.~(\ref{eq:eigenvalue_equation}) results in a more symmetric form \cite{nigg2012}:
\begin{align}
\label{eq:symmetrized_eigenvalue}
 \boldsymbol{\Omega}^2\mathbf{u}_{\mu} =~& \omega^2_{\mu} \mathbf{u}_{\mu}, \\
 \boldsymbol{\Omega}^2 \equiv~& \mathbf{\widetilde C}^{-1/2}\mathbf{\widetilde L}^{-1}\mathbf{\widetilde C}^{-1/2},
\label{eq:Omega_squared}
\end{align}
where $\mathbf{u}_{\mu}$ represents the transformed eigenvector, and the matrix $\bm{\Omega}^2$ is symmetric by definition. 
This symmetrization strategy allows us to factor $\mathbf{\widetilde C}^{1/2}$ out of the impedance matrix $\mathbf{Z}^{(0)}(\omega)$ and rewrite it strictly in terms of its poles:
\begin{align}
&\mathbf{Z}^{(0)}\left(\omega\right)=\mathbf{\widetilde C}^{-1/2}\frac{1}{\left(j\omega\right)^{-1}\boldsymbol{\Omega}^{2}+j\omega}\mathbf{\widetilde C}^{-1/2} \nonumber \\
&=\sum_{\mu=1}^N\left(\mathbf{\widetilde C}^{-1/2}\mathbf{e}_{\mu}\right)\frac{1}{\left(j\omega\right)^{-1}\omega^{2}_{\mu}+j\omega}\left(\mathbf{\widetilde C}^{-1/2}\mathbf{e}_{\mu}\right)^{T}\nonumber \\
&=\sum_{\mu=1}^N\frac{2j\omega\omega_{\mu}}{\omega^{2}_{\mu}-\omega^{2}}\left(\frac{\mathbf{v}_{\mu}}{2\sqrt{\omega_{\mu}\mathcal{E}_{\mu}}}\right)\left(\frac{\mathbf{v}_{\mu}}{2\sqrt{\omega_{\mu}\mathcal{E}_{\mu}}}\right)^{T},
\label{eq:z_summation_proof}
\end{align}
where in the second line we introduced the normalized eigenvectors $\mathbf{e}_{\mu} = \mathbf{\widetilde C}^{1/2}\mathbf{v}_{\mu} / |\mathbf{\widetilde C}^{1/2}\mathbf{v}_{\mu}|$, and in the third line we introduced mode energy $\mathcal{E}_{\mu} = |\mathbf{\widetilde C}^{1/2}\mathbf{v}_{\mu}|^2/2$.

Equation~(\ref{eq:z_summation_proof}) is the primary result of this section. It proves the results in Eqs.~(\ref{eq:sum_over_poles},\ref{eq:residue_matrix}) that the impedance matrix $\mathbf{Z}^{(0)}(\omega)$ can indeed be written in the form of Eq.~(\ref{eq:sum_over_poles}), and the residue matrix $\mathbf R_{\mu}$ can be written as a product of a column vector and its transpose. Comparing Eq.~(\ref{eq:z_summation_proof}) with Eqs.~(\ref{eq:sum_over_poles},\ref{eq:residue_matrix}), we find the expression for $\boldsymbol{\sigma}_{\mu}$ in terms of the voltage eigenvectors to be
\begin{equation}
\boldsymbol{\sigma}_{\mu}=\frac{\mathbf{v}_{\mu}}{2\sqrt{\omega_{\mu}\mathcal{E}_{\mu}}}
\end{equation}
and the matrix $\boldsymbol{\Sigma}$ defined in Eq.~(\ref{eq:sigma_matrix_stack}) to be
\begin{equation}
\boldsymbol{\Sigma}=\frac{1}{2}\sum_{\mu=1}^N\frac{\mathbf{v}_{\mu}\mathbf{e}^{T}_{\mu}}{\sqrt{\omega_{\mu}\mathcal{E}_{\mu}}}
\label{eq:sigma_result}
\end{equation}
Because $\mathbf{v}_{\mu}$ scales proportionally to $\sqrt{\mathcal{E}_{\mu}}$, the matrix $\boldsymbol{\Sigma}$ is independent of $\sqrt{\mathcal{E}_{\mu}}$. The results in Eq.~(\ref{eq:sigma_to_lc}) follow from Eqs.~(\ref{eq:Omega_squared}, \ref{eq:sigma_result}).

The matrix $\boldsymbol{\Sigma}$ carries a clear physical meaning. 
The matrix element $\varSigma_{n\mu}$ multiplied by $\sqrt{\hbar}$ is equal to the zero-point fluctuations of flux at node $n$ created by mode $\mu$. It is easy to verify this in the case of a single LC-oscillator (i.e., $N = 1$). In this case, we have from Eq.~(\ref{eq:sigma_result}) that $\mathbf\Sigma =\sqrt{\hbar/2}(\widetilde L/\widetilde C)^{1/4}$ matching the familiar expression of zero point flux fluctuations of a single LC oscillator. 

Finally, we elaborate on the origin of $\widetilde{\delta\mathbf{L}}$ term in Eq.~(\ref{eq:Z_convenient}). Any physical system contains a large number of high-frequency modes associated with the smaller elements in the layout. In the qubit-cavity system we study, they can come from modes associated with the geometric wire inductances of qubits and high frequency cavity modes. While these modes sit outside our simulated frequency band, the low-frequency tails of their driven response still alter our simulated impedance matrix. As can be seen from Eq.~(\ref{eq:z_summation_proof}), the low-frequency tails of impedance response of high-frequency modes scale linearly with frequency $\omega$ to the leading order, and thus behaves like effective inductances. This is the physical origin of the $\widetilde{\delta\mathbf{L}}$ term in Eq.~(\ref{eq:Z_convenient}).

% \color{teal}

\section{Longitudinal section magnetic (LSM) package modes}
\label{appendix:LSM_modes_derivation}

\begin{figure}[htbp]
    \centering
    % Replace 'my_figure.pdf' with your actual PDF filename
    \includegraphics[trim=6.1cm 3.6cm 6cm 2.5cm, clip, width=0.9\linewidth]{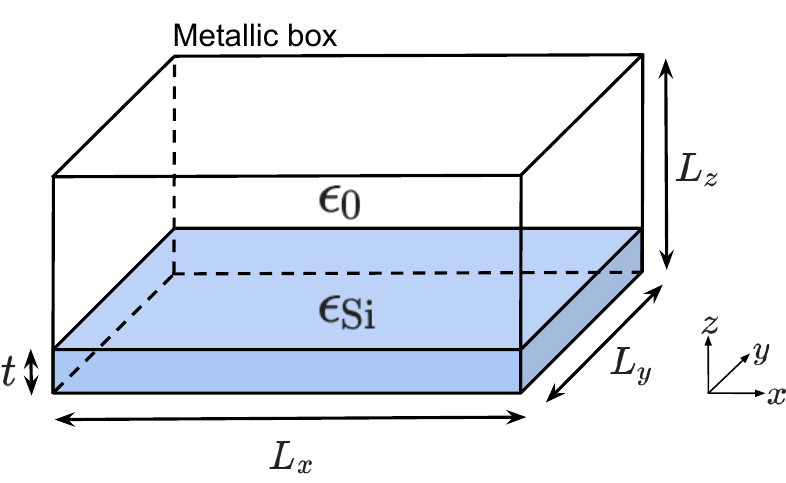} 
    \caption{Metallic package box partially filled with dielectric.}
    \label{fig:Appendix_E_fig} % Used for cross-referencing later
\end{figure}

In this section we discuss the electromagnetic fields of  Longitudinal-Section-Magnetic~\cite{balanis2012} (LSM) modes that occur in rectangular resonant cavities with a dielectric slab (without qubit paddles), as shown in Fig.~\ref{fig:Appendix_E_fig}. This figure is a simplified version of Fig.~\ref{fig:hfss} of the main text.

The LSM modes are eigensolutions of Maxwell's equations with a vanishing vertical magnetic field (i.e., $H_z=0$ everywhere). These eigenmodes are indexed by three integers; namely, $a,b>0$ and $\nu\geq0$. These indices parametrize the eigenmode frequency, $\omega_{ab\nu}$. 

\begin{figure}[t!]
    \centering
    \includegraphics[trim=3cm .5cm 3cm 0cm, clip, width=0.95\linewidth]{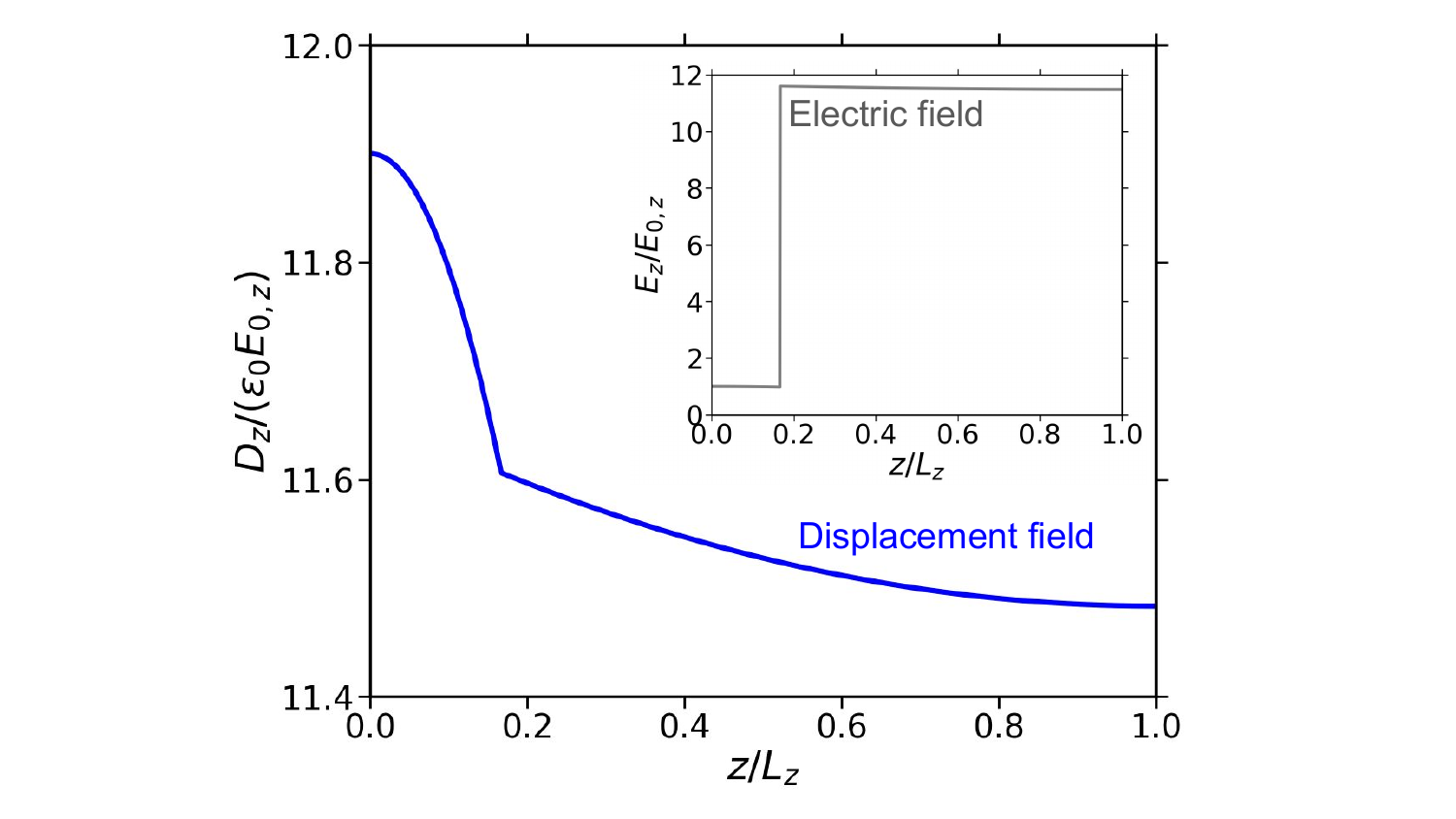} 
    \caption{Vertical components of the displacement and electric fields evaluated at $x=L_x/2$ and $y=L_y/2$ for the fundamental mode. Both fields vary weakly away from the dielectric-air interface ($z=t$); the electric field is discontinuous at $z=t$. We use geometric parameters given in the main text and dielectric relative permittivity of $\varepsilon_r=11.9$. }
    \label{fig:Appendix_E_fig_Ez} % Used for cross-referencing later
\end{figure}

For the eigenmode $\text{LSM}_{ab\nu}$, the vertical component of the electric field in the dielectric region ($0\leq z\leq t$) is  
\begin{align}
    E_z^{\rm Si} =&\, {E}_{0,z}\sin\left(\frac{a\pi}{L_x}x\right)\sin\left(\frac{b\pi}{L_y}y\right)\cos\left(k_{z}^{\rm Si}\, z\right),\label{eq:LSM_Ez_Si}
\end{align}
while in the air region ($t\leq z\leq L_z$) is given by 
\begin{align}
    E_z^{\rm air} =&\, \varepsilon_r{E}_{0,z} \frac{\cos(k_z^{\rm Si}t)}{\cosh(k_z^{\rm air}(L_z-t))} \sin\left(\frac{a\pi}{L_x}x\right)\sin\left(\frac{b\pi}{L_y}y\right)\nonumber\\
    &\,\times \cosh\left(k_{z}^{\rm air}\, (L_z-z)\right).\label{eq:LSM_Ez_air}
\end{align}
Note that the above formulas satisfy the condition that the vertical component of the displacement field is continuous across the dielectric-air interface (see Fig.~\ref{fig:Appendix_E_fig_Ez}). 

The parameters $k_z^{\rm Si}$ and $k_z^{\rm air}$ together with the mode eigenfrequency $\omega_{ab\nu}$ are obtained from the solution of the following transcendental equation, 
\begin{align}
    k_{z}^{\rm Si}\tan(k_{z}^{\rm Si} t) =& \,\varepsilon_r k_{z}^{\rm air} \tanh\big(k_{z}^{\rm air}(L_z-t)\big),\label{eq:transcendental_eq}
\end{align}
where
\begin{align}
    k_z^{\rm Si} =&\, \sqrt{\varepsilon_{r} (\omega_{ab\nu}/c_0)^2 - \gamma_{ab}^2},\label{eq:kz_Si_eq}\\
    k_z^{\rm air} =&\, \sqrt{\gamma_{ab}^2 - (\omega_{ab\nu}/c_0)^2},\label{eq:kz_air_eq} \\    
    \gamma_{ab}=&\,\sqrt{\left(\frac{a\pi}{L_x} \right)^2 + \left(\frac{b\pi}{L_y} \right)^2},\label{eq:gamma_ab}    
\end{align}
$c_0$ is the speed of light in the air region and $\varepsilon_r$ is the relative permittivity of the dielectric slab. For given $a$ and $b$, there are several real solutions for $\omega_{ab\nu}$ indicated by the index $\nu\geq0$. The smallest solution, denoted by $\nu=0$, corresponds to the fundamental mode ${\rm LSM}_{110}$. For this mode, both  $k_z^{\rm Si}$ and $k_{z}^{\rm air}$ are real and small  which implies a weak $z$-dependence of the electric field, albeit with a jump at the dielectric-air interface, see inset of Fig.~\ref{fig:Appendix_E_fig_Ez}. Note that $\partial_z E_z$ is continuous at the interface, which is necessary for the continuity of the $x$ and $y$ components of the electric field at the interface, as seen from Eq.~\eqref{eq:LSM_Ex_Ey} below. This condition together with the continuity of the vertical component of the displacement field lead to the above transcendental equation~\eqref{eq:transcendental_eq}. 

The other components of the electric field are related to $E_z$ in both dielectric and air regions as 
\begin{align}
    E_x = \gamma_{ab}^{-2}\partial^2_{xz}E_z \;\; {\rm and}\;\; E_y = \gamma_{ab}^{-2}\partial^2_{yz}E_z.  \label{eq:LSM_Ex_Ey}
\end{align}
Note that $E_x$ and $E_y$ vanish at the top and bottom of the metallic box due to perfect electric conductor boundary conditions at the box walls.

The $x$ component of the electric field in the plane of the qubits ($z=t$) is given by 
\begin{equation}
E_x(x,y,z=t) = a E_{0,x}\cos\left(\frac{a\pi}{L_x}x\right)\sin\left(\frac{b\pi}{L_y}y\right), 
\label{eq:Ex_Si_qubits_plane}
\end{equation}
where 
\begin{align}
    E_{0,x} = -{E}_{0,z}\frac{(\pi/L_x)\, k_{z}^{\rm Si}\sin\left(k_{z}^{\rm Si}\,t\right)}{\gamma_{ab}^2}. \label{eq:E0_formula}
\end{align}
\newline 
The magnetic field is related to $E_z$ as follows
\begin{align}
    H_x &= -i\epsilon\,\omega_{ab\nu}\,\gamma_{ab}^{-2}\partial_y E_z, \label{eq:LSM_Hx}\\
    H_y &= i\epsilon\,\omega_{ab\nu}\,\gamma_{ab}^{-2}\partial_x E_z,\label{eq:LSM_Hy}\\ 
    H_z &= 0,\label{eq:LSM_Hz} 
\end{align}
where $\epsilon$ is the permittivity at the dielectric ($\epsilon=\epsilon_{\rm Si}$) or air regions ($\epsilon=\epsilon_0$). 

Note that in the above expressions we do not include the time dependence of the fields. To include it, we multiply these expressions by the phase factor $\exp(-i\omega_{ab\nu} t)$, with $t$ being time, and then take the real part.

In the main text we normalize the electric field such that the total energy of the $\text{LSM}$ mode is $\mathcal{E}_m$. To do this normalization, we first calculate the mode energy $U_{\rm LSM}=\int_{\rm Box}\text{d}v\, \epsilon(\mathbf{r})|\vec E(\mathbf{r})|^2/2$, which is given by 
\begin{align}
 \hspace{-0.1cm}   \frac{U_{\rm LSM}}{U_0} =&\, \varepsilon_r\Bigg[\Big(1 + \big(\frac{k_z^{\rm Si}}{\gamma_{ab}}\big)^2\Big)\frac{t}{L_z} + \Big( 1 - \big(\frac{k_z^{\rm Si}}{\gamma_{ab}}\big)^2\Big)\frac{\sin(2k_z^{\rm Si}t)}{2k_z^{\rm Si}L_z}\Bigg]\nonumber\\
    &+ \frac{\varepsilon_r^2\cos^2(k_z^{\rm Si}t)}{\cosh^2(k_z^{\rm air}(L_z-t))}\Bigg[\Big(1 - \big(\frac{k_z^{\rm air}}{\gamma_{ab}}\big)^2 \Big)\big(1-\frac{t}{L_z}\big) \nonumber\\ 
    & + \Big(1 + \big(\frac{k_z^{\rm air}}{\gamma_{ab}}\big)^2\Big)\frac{\sinh(2k_z^{\rm air}(L_z-t))}{2k_z^{\rm air}L_z}\Bigg].
    \label{eq:U_LSM}
\end{align}
Here $U_0 = \epsilon_0{E}_{0,z}^2 \mathcal{V}/16$ and  $\mathcal{V}$ is the box volume. Then, using Eqs.~\eqref{eq:E0_formula} and~\eqref{eq:U_LSM}, we find the prefactor $E_{0,x}$ in Eq.~\eqref{eq:Ex_LSM-t} of the main text,
\begin{align}
    E_{0,x} =& \frac{(\pi/L_x) k_z^{\rm Si}\sin(k_z^{\rm Si}t)}{\gamma_{ab}^2} \sqrt{\frac{\mathcal{E}_m}{(\epsilon_0\mathcal{V}/16)(U_{\rm LSM}/U_0)}}.
    \label{eq:E0_expression}
\end{align}
The above formula is valid for any $\text{LSM}$ mode ($\nu \geq0$). 

For the fundamental package mode ($\text{LSM}_{110}$) and parameters $L_x=L_y=30$~mm, $L_z=3$~mm, $t=0.5$~mm and $\varepsilon_r=11.9$, we obtain the fundamental mode frequency $\omega_{110}=2\pi\times 6.49$~GHz from the solution of the above transcendental equation, and $k_z^{\rm Si}= 445.4\,{\rm m}^{-1}$ and $k_z^{\rm air}= 58.4\,{\rm m}^{-1}$. Note that  $2\pi/k_z^{\rm Si}$ and $1/k_z^{\rm air}$ are much larger than the thickness of the dielectric slab and the 2.5~mm thickness of the air region, respectively, which implies that the fields of the fundamental mode vary weakly along the vertical direction, see Fig.~\ref{fig:Appendix_E_fig_Ez}. From Eqs.~\eqref{eq:U_LSM}--\eqref{eq:E0_expression}, we obtain the prefactor $E_{0,x}$ in Eq.~\eqref{eq:Ex_LSM-t} of the main text; we find $E_{0,x}=25.56\times10^6$~V/m (assuming a mode energy of $\mathcal{E}_m=1\,{\rm J}$). 

If we assume $k_z^{\rm Si}t$ and $k_z^{\rm air}(L_z-t)$ are small for $\text{LSM}_{ab0}$ modes (e.g., for the fundamental mode, they are approximately 0.22 and 0.15, respectively), we can approximately solve the transcendental equation~\eqref{eq:transcendental_eq} using the approximation $\tan \delta\approx \delta$, for $|\delta|\ll1$. This leads to an approximate formula for the mode frequency; namely, $\omega_{ab0} \approx c_0\gamma_{ab}/\sqrt{\varepsilon_{r;\, \rm eff}} $, where $\varepsilon_{r;\,\rm eff}\equiv [\varepsilon_r^{-1}(t/L_z) + (1-t/L_z)]^{-1}$ is the effective relative permittivity for the cavity. This simple formula is rather accurate; for instance, for the fundamental mode ($a=b=1$), it gives a mode frequency of 6.50~GHz that is only $\sim 0.2\%$ off from the exact value of 6.49~GHz. An approximate expression for $k_z^{\rm Si}$ is $k_z^{\rm Si}\approx  \gamma_{ab} \sqrt{(\varepsilon_r-1)(1-t/L_z)}$, which, for the fundamental mode, gives $446.3\,{\rm m}^{-1}$ while the exact value is $445.4\,{\rm m}^{-1}$. Similarly, we find $k_z^{\rm air}\approx \gamma_{ab}\sqrt{1-\varepsilon_{r;\,\rm eff}^{-1}}$, which, for the fundamental mode, gives $57.9\,{\rm m}^{-1}$ that is again close to the exact value of $58.4\,{\rm m}^{-1}$. Moreover, the mode energy $U_{\rm LSM}$ can be approximated as $U_{\rm LSM}\approx [\varepsilon_r(t/L_z) + \varepsilon_r^2(1-t/L_z)] \epsilon_0{E}_{0,z}^2 \mathcal{V}/8$. Using these approximate results, the normalization coefficient $E_{0,x}$ of Eq.~\eqref{eq:Ex_LSM-t} of the main text can be written as ($\mathcal{V}=L_xL_yL_z$)
\begin{align}
    E_{0,x}\approx&\, (\varepsilon_r-1)\frac{\pi t}{L_x}(1-\frac{t}{L_z}) \sqrt{\frac{8 \mathcal{E}_m}{\epsilon_{\rm Si}\, \mathcal{V}}\frac{\varepsilon_{r;\,{\rm eff}}}{\varepsilon_r}}.
    \label{eq:E0_approx}
\end{align}
For the fundamental mode, the above formula gives $25.1\times10^6$~V/m while the exact result is $25.56\times10^6$~V/m. Note  that the above estimation of $E_{0,x}$ is independent of mode parameters $a$ and $b$. Indeed, for the mode $\text{LSM}_{210}$ we obtain $E_{0,x}=26.25\times10^6\,{\rm V}/{\rm m}$, using the exact formula~\eqref{eq:E0_expression} and assuming a mode energy of $\mathcal{E}_m=1\,{\rm J}$, which is rather close to the value of $E_{0,x}$ of the fundamental mode.

\renewcommand{\bibsection}{
\end{document}